%% file: ms.tex
\documentclass[10pt,preprint]{aastex}
\tighten 
\begin{document}
\newcommand{\lya}{Lyman~$\alpha$}
\newcommand{\lyb}{Lyman~$\beta$}
\newcommand{\za}{$z_{\rm abs}$}
\newcommand{\ze}{$z_{\rm em}$}
\newcommand{\cmtwo}{cm$^{-2}$}
\newcommand{\nhi}{$N$(H$^0$)}
\newcommand{\degpoint}{\mbox{$^\circ\mskip-7.0mu.\,$}}
\newcommand{\kms}{\,km~s$^{-1}$}      
\newcommand{\minpoint}{\mbox{$'\mskip-4.7mu.\mskip0.8mu$}}
\newcommand{\peryr}{\mbox{$\>\rm yr^{-1}$}}
\newcommand{\secpoint}{\mbox{$''\mskip-7.6mu.\,$}}
\newcommand{\sqdeg}{\mbox{${\rm deg}^2$}}
\newcommand{\squig}{\sim\!\!}
\newcommand{\subsun}{\mbox{$_{\twelvesy\odot}$}}
\newcommand{\et}{{\rm et al.}~}

\def\ltsima{$\; \buildrel < \over \sim \;$}
\def\simlt{\lower.5ex\hbox{\ltsima}}
\def\gtsima{$\; \buildrel > \over \sim \;$}
\def\simgt{\lower.5ex\hbox{\gtsima}}
\def\arcs{$''~$}
\def\arcm{$'~$}
\def\erf{\mathop{\rm erf}}
\def\erfc{\mathop{\rm erfc}}
\title{LYMAN BREAK GALAXIES AT REDSHIFT Z$\sim$3: SURVEY DESCRIPTION AND FULL
DATA SET\altaffilmark{1} }
\author{\sc Charles C. Steidel\altaffilmark{2} }
\affil{California Institute of Technology, MS 105--24, Pasadena, CA 91125}
\author{\sc Kurt L. Adelberger\altaffilmark{3}}
\affil{Harvard-Smithsonian Center for Astrophysics, 60 Garden St., Cambridge, MA 02139}
\author{\sc Alice E. Shapley}
\affil{California Institute of Technology, MS 105--24, Pasadena, CA 91125}
\author{\sc Max Pettini}
\affil{Institute of Astronomy, Madingley Road, Cambridge CB3 OHA, UK}
\author{\sc Mark Dickinson and Mauro Giavalisco}
\affil{Space Telescope Science Institute, 3700 San Martin Drive, Baltimore, MD 21218}


\altaffiltext{1}{Based, in part, on data obtained at the 
W.M. Keck Observatory, which 
is operated as a scientific partnership among the California Institute of Technology, the
University of California, and NASA, and was made possible by the generous financial
support of the W.M. Keck Foundation.
} 
\altaffiltext{2}{Packard Fellow}
\altaffiltext{3}{Harvard Society Junior Fellow}
\begin{abstract}
We present the basic data for a large ground-based spectroscopic 
survey for $z \sim 3$ ``Lyman break galaxies'' (LBGs),
photometrically selected using rest-UV colors from very deep images in 17 high Galactic latitude fields.
The total survey covers an area of 0.38 square degrees, and includes 2347 photometrically-selected
candidate LBGs to an
apparent ${\cal R}_{\rm AB}$ magnitude limit of 25.5. Approximately half of these objects
have been observed spectroscopically using the Keck telescopes, yielding
940 redshifts with $\langle z\rangle =2.96\pm0.29$. 
We discuss the images, photometry, target selection, and the spectroscopic 
program in some detail, and present catalogs of the photometric and spectroscopic
data, made available in electronic form. We discuss the general utility 
of conducting nearly-volume-limited redshift surveys in prescribed redshift
intervals using judicious application of photometric pre-selection.    
\end{abstract}

\section{INTRODUCTION}

The advent of 10-m class telescopes in the mid-1990s provided for the first time the capability
of relatively routine spectroscopic observations of galaxies at very high
redshifts (e.g., Steidel \et 1996a). However, as was well-known from the most ambitious apparent-magnitude
selected surveys conducted on 4m-class telescopes (e.g., Lilly \et 1996; Ellis \et 1996) and
early work on the Keck telescopes (e.g., Songaila \et 1994, Cohen \et 1996, Cowie \et 1996), simply observing
fainter galaxies is a relatively inefficient means of assembling significant statistical samples
at high redshift, and even at the limit of 10-m class telescopes for complete spectroscopic
samples (R$\sim 24$) the sampling beyond $z \sim 1$ is relatively sparse.  
Meanwhile, the technique of ``photometric redshifts'' had gained considerable new 
impetus as imaging surveys such as the Hubble Deep Field campaigns (Williams \et 1996)
reach far deeper than current spectroscopic capabilities, and some insights into the
galaxy formation process are possible with redshift estimates only. 

Our approach has been based on the assertion that the information gained from spectroscopy is essential
for addressing many scientific issues. The technique for acquiring the spectroscopic information
might be thought of as a hybrid that lies somewhere between photometric redshifts and
traditional flux-limited ``blind'' redshift surveys. ``Photometric pre-selection'',
where objects are chosen for spectroscopy based on coarse features in their spectral
energy distributions that can be recognized
from simple broad or intermediate-band photometry, introduces both
advantages and complications (see, e.g., Steidel 2000, Adelberger 2002). The technique
has several practical advantages: for example, one can
focus on a particular range of redshift using appropriately chosen color criteria, 
so that the imposition of an apparent magnitude limit 
is nearly equivalent to imposing a luminosity limit. If the selection function
is relatively narrow in redshift, one can construct nearly volume-limited samples, so that one can include
a significant range in absolute magnitude at a given redshift without observing many objects
outside of the targeted redshift range. 
Photometric pre-selection mitigates many issues 
inherent in flux limited surveys, such as very broad redshift selection functions where different
kinds of objects are selected at the ``front'' and the ``back'' of the survey volume. 
There is also a distinctly practical advantage of efficiency for spectroscopy on large telescopes-- at an apparent
magnitude of $R \sim 25$, the surface density of faint galaxies has reached $\sim 30$ per
square arc minute, much too dense for complete spectroscopy without observing the same patch
of sky many times using different slit masks. 
The redshift distribution of such a sample would be extremely broad ($z \sim 0.1-5$), and typical
spectroscopic observations optimized over less than an octave in wavelength would fail to
measure redshifts for a large fraction\footnote{For example, one would want to be optimized in the
$7000-10000$ \AA\ range for $z \sim 0.9-1.5$, in the 4000-6000 \AA\ range for $z \simgt 2.5$, 
and in either the near-UV or the near-IR for redshifts in between.} of the targeted galaxies. 
The gain in efficiency for studying galaxies in a particular constrained cosmological
volume can be as large as a factor of 10--25, depending on the chosen redshift range and nature
of the selection on coarse spectral properties. 
With photometric pre-selection, one knows {\it a priori} the 
approximate redshift of the targets and thus
the optimal spectroscopic configuration (wavelength range, grating and blaze wavelength, 
spectral resolution required) that balances efficiency and information content.
Disadvantages of photometric pre-selection include increased complexity of
the selection function, the need for higher quality initial images for target selection, and
more difficult post-facto analysis to understand issues of completeness. Many of these
issues are discussed in some detail in Steidel \et (1999) and Adelberger (2002). 

The use of the Lyman limit of hydrogen at 912 \AA\ in the rest frame of the galaxies
is only one example of all potential uses of photometric pre-selection; more recently,
Adelberger (2000) and Davis \et (2002) have applied photometric culling for selecting samples
with $z=1.0\pm0.1$ and $z \simgt 0.7$, respectively. The many on-going searches for 
``Extremely Red Objects'' selected by virtue of relatively extreme optical-IR colors are similar
in spirit (e.g., McCarthy \et 2001; Daddi \et 2002), as are searches targeting high-redshift objects selected by
their IR colors(e.g., Labb\'e et al. 2003). Targeting $z \sim 3$ for our initial survey
grew out of purely practical issues: it is the lowest redshift where the Lyman limit
can be recognized from the ground using broad-band photometry (Guhathakurta, Tyson, \& Majewski 1990; 
Steidel \& Hamilton 1992, 1993) because the standard U band (or, $U_n$, in our case) is centered
near 3550 \AA.  

The use of broad-band colors to select objects for subsequent spectroscopy has long been 
a standard technique to search for quasars (e.g., Green 1976; Koo \& Kron 1988; Warren, Hewett, \& Osmer 1991), and the suggestion of using
colors to identify primeval galaxies was made as early as 25 years ago (Meier 1976), 
with searches commencing soon after (e.g., Koo \& Kron 1980). However, most of the early searches for
high redshift galaxies depended on an assumed extremely luminous phase expected if massive
galaxies formed a large fraction of their stars on a dynamical timescale (motivated
by arguments given in, e.g., Eggen, Lynden-Bell, \& Sandage 1962 and Partridge \& Peebles 1967) and
in any case were limited by the relatively poor sensitivity of photographic plates. More
stringent constraints on the {\it existence} of star-forming galaxies at redshifts
significantly beyond $z \sim 1$ had to await the first
extremely deep CCD imaging surveys. Meanwhile, we were strongly influenced by the results  
of QSO absorption line studies (e.g. Wolfe \et 1986;
Sargent \et 1988, 1989; Steidel 1990a,b, Bergeron \& Boiss\'e 1991), some of which suggested that relatively
normal galaxies may be present at redshifts as high as $z=2-3$. 

The general lack of success of
most ``proto-galaxy'' surveys (see, e.g.,  the review by Pritchet 1994) that focused on
the detection of the Lyman $\alpha$ emission line suggested that other techniques using features
less susceptible to dust and scattering effects were necessary before high redshift
galaxies were written off as non-existent, hopelessly faint,  or completely hidden. 
Most of the emphasis in the galaxy evolution community
through the late 1980s and early 1990s
was on the so-called ``faint blue galaxies'' which became especially evident in very deep CCD
images of the sky (e.g., Tyson 1988; Cowie \et 1988; Lilly \et 1991). The fact that the galaxies
were very blue from $B$ to $K$ (Cowie \et 1988) implied that these objects must lie
at redshifts $z \le 3.5$. 
Guhathakurta, Tyson, \& Majewski (1990) set more explicit limits on the fraction of ``faint blue galaxies''  
to apparent magnitude limits of $R \sim 26$ that could lie at redshifts $z \simgt 3$, based on the observed flatness of the spectral energy
distributions through the observed $U$ band. They concluded that the fraction of galaxies
with $z \simgt 3$ was less 7\%, and they suspected
that it ``is likely to be less than 1\%-2\%''\footnote{In fact, with hindsight, the fraction of
galaxies to ${\cal R}=26$ that satisfy our LBG selection criteria is 6.3\%, from our survey in the Q1422+2309
field. This apparently small number
says more about the high density of ``foreground'' objects than about the
absence of high redshift objects, and emphasizes why photometric pre-selection is essential
for targeting a particular range of redshifts. }. 
Steidel \& Hamilton (1992, 1993) and Steidel, Pettini, \& Hamilton
(1995) addressed a somewhat different follow-up question: given knowledge of where to look for a galaxy
with a redshift $z \ge 3$, can one either detect it or set limits on its luminosity using a technique
tuned to find even intrinsically faint objects at the known redshift of Lyman limit systems
or damped Lyman $\alpha$ systems, even if they have no Lyman $\alpha$ emission?   
Some success in identifying {\it candidate} high redshift galaxies resulted from these
searches targeting ``known'' galaxies near the lines of sight to background QSOs.
The first spectroscopy of these candidates was however not attempted until the fall
of 1995 (Steidel \et 1996), using the Keck 10m telescope and the 
Low Resolution Imaging Spectrometer (LRIS; Oke \et 1995).
Unexpected success in measuring spectroscopic redshifts with reasonable integration times
suggested that more comprehensive surveys for high redshift galaxies were feasible.  
This paper represents a superset of the observations made after converging on a uniform
approach to candidate selection and spectroscopy. 

A number of other practical advantages of working near $z \sim 3$ became clear from the initial 
spectra: Lyman $\alpha$ emission is not required to confirm redshifts, because a large number 
of strong interstellar lines in the rest wavelength range 1200-1700 \AA\ falls conveniently
in the 4500-6500 \AA\ observed range, where optical spectrographs are efficient and 
the sky background 
is relatively dark. Thus, quite unexpectedly, spectroscopic identification with a very
high success rate was possible to apparent magnitudes ${\cal R}\sim 25.5-26$, as much as
two magnitudes fainter than the reasonable limit for a simple apparent magnitude selected
survey that would yield a similar spectroscopic success rate. Of course, there are certainly
objects in the targeted redshift range that would not satisfy our color selection criteria
and so could not be accounted for even with careful correction for incompleteness (e.g., objects
that are reddened with more than $E(B-V) \sim 0.4$, corresponding to $\sim 4-5$ magnitudes of
internal extinction in the far-UV -- Adelberger \& Steidel 2000). Other techniques, such as
sub-mm or X-ray selection, are required to find such objects. However, by number, objects
that satisfy the simple LBG color selection rules are by far the dominant ``population''
at $z \sim 3$ (Dickinson 2000). Whether they dominate the total star formation rate or energy production
rate is still a matter of significant debate (cf. Adelberger \& Steidel 2000; Barger, Cowie,
\& Sanders 1999). It is certainly safest to treat LBGs as a type of object that is common,
easy to find using readily available observational facilities, and for which large statistical
samples are possible. No doubt our understanding of how LBGs are related to other objects
in the $z \sim 3$ universe will improve substantially in the coming years.  
 
The data collected in this paper have been used in a number of more focused investigations
on the nature of Lyman break galaxies, their large-scale distribution, their relationship to the
diffuse intergalactic medium, and their connection with other high redshift samples of galaxies
selected at other wavelengths. For example, subsets of the spectroscopic sample have been
used to quantify the clustering strength of $z \sim 3$ LBGs (Steidel \et 1998; Adelberger \et 1998;
Adelberger \et 2003)
while the photometric samples, together with knowledge of the redshift selection function, have
been used to measure clustering from various angular statistics (Giavalisco \et 1998; Giavalisco \& Dickinson
2001; Porciani \& Giavalisco 2002). The redshift distribution of a substantial subset
of the sample presented here was used, together with the photometric sample,
to evaluate the rest-frame
far--UV luminosity function (Dickinson 1998; Steidel \et 1999; Adelberger \& Steidel 2000),
the range of intrinsic properties represented in the LBG sample, and the likely contribution
of UV-selected objects to star formation history and far--IR/sub-mm background. Studies of the 
near-IR spectra (Pettini \et 1998, 2001)
and near-IR photometry (Shapley \et 2001) have made use of subsets of the spectroscopic sample presented
here, to explore kinematics, chemical abundances, and star-formation histories. The trends discernible
in the optical discovery spectra of the LBGs have been explored and quantified by Shapley \et (2003).
The AGN content of the spectroscopic sample has been discussed in a preliminary way by Steidel \et (2002). 
Some of the LBGs presented here have been examined at X-ray (Nandra \et 2002) and sub-mm (Chapman \et 2000;
Chapman \et 2001) wavelengths. Almost half of the current sample of $z\sim 3$ LBGs was obtained as  
part of a major effort to  quantify the relationship between H~I and metals in the intergalactic
medium and the star forming galaxies (Adelberger \et 2003), and has not been used in many of
the aforementioned studies because it was not completed until 2002. 

It is not our intention to duplicate any of the work that has already been published on the $z\sim 3$
LBG sample, but rather to present the photometry and spectroscopy for the entire sample
in as compact a manner as possible. We describe in more detail the overall survey strategy, from
a practical point of view, and present some statistics that may prove useful to other workers
who might like to make use of some or all of the data. The organization of the paper is
as follows: \S 2 describes the survey imaging data and reductions; \S 3 summarizes the 
photometry; \S 4 describes the photometric selection of candidate LBGs; \S 5 discusses
the spectroscopic observations and analysis of the candidates; \S 6 summarizes the
results for ``interloper'' objects in the spectroscopic sample, and \S 7 includes discussions
of individual fields in the survey. A short summary of the results is given in \S 8.

\section{IMAGING OBSERVATIONS AND DATA}

The imaging data for the 17 fields presented here were obtained during the
interval 1994--2000 on a number of telescopes, detailed in Table 1. 
The imaging filters used were the $U_nG{\cal R}$ system (supplemented with
$I$ band data in most cases, although the $I$ band data were not explicitly
used for the $z \sim 3$ LBG project) described in Steidel \& Hamilton (1993);
it turns out that the $U_n$ filter is essentially identical to the 
Sloan Digital Sky Survey (SDSS) u' filter (both $U_n$ and SDSS u' cut at
4000 \AA\ on the red side, about 200 \AA\ blue-ward of the standard Johnson U
passband). The $G$ filter has the same effective wavelength as the SDSS g' filter,
but is slightly narrower. The ${\cal R}$ filter used for the vast majority
of the imaging presented in this paper has a bandpass centered at 6830 \AA\ and
a half-power bandwidth of $\sim 1250$ \AA. The filter system passbands are shown in
Figure 1.   

In general, individual integrations of 900s (${\cal R}$), 1200s ($G$) and 
1800s ($U_n$) were obtained, with 10-15\arcsec\ dithers between exposures.
Wherever possible, the $U_n$ data were obtained within $\sim 2$ hours of the
meridian to minimize atmospheric attenuation and dispersion effects.
For all but some of the WHT data, the detectors used were thinned, back-side illuminated
Tektronix (SITe) 2048 x 2048 CCDs, with a pixel scale depending on the telescope
(see Table 1). The data for Q1422+2309 and Q0302-003 were obtained at the William Herschel
Telescope using, respectively, a single 2K x 4K EEV (now Marconi) CCD
and a mosaic of 2 such EEV devices. These CCDs have approximately twice the UV quantum
efficiency of the SITe devices but suffer from much higher amplitude fringing in the
red. With very few exceptions, data obtained under seeing condition worse than
1.3\arcsec\ (FWHM) were excluded, so that the final stacked images average
$\sim 1$\arcsec\ (Table 1). 
Total integration times varied depending on observing conditions and 
telescope/camera combinations; these are also summarized in table 1. 

The data were flat-fielded using dome flats for $G$, ${\cal R}$, and $I$ band 
observations, and using a combination of twilight sky and dark sky ``superflats'' (where
flat fields were created by median-combining dis-registered science images after masking out the
positions of objects) for
the $U_n$ band data. Fringe removal for the $I$ band images (and ${\cal R}$ for the WHT EEV
data) was accomplished using standard techniques. Cosmic rays were identified 
morphologically on each exposure and masked so as to be excluded in the production of the
stacked image. Stacks were produced by combining sub-pixel-registered, photometrically-scaled
images in each band and averaging unmasked pixels. Final stacks in each band were
then geometrically transformed onto the pixel scale of the ${\cal R}$ image so that
the final images in each band are registered to better than 0.05\arcsec. 
To avoid photometric non-uniformities in the survey fields, for all single-CCD data
we retained only those image regions receiving the full exposure time in every band. 

Several spectrophotometric standard stars from Massey \et (1988) or from Oke (1990) 
were observed during each observing night, through each filter, at a range
of airmasses. Artificial magnitudes on the AB system
for $U_n$, $G$, ${\cal R}$, and $I$ for each standard were produced by multiplying 
the effective filter passband (including CCD quantum efficiency and atmospheric 
attenuation) and the stellar energy distribution. Photometric zero points (reduced
to an airmass of 1.0) measured in this way had internal scatter of $\pm 0.02$ magnitudes
for $G$, ${\cal R}$, and $I$ filters, and $\pm 0.03$ magnitudes for $U_n$ among different
standard star measurements obtained during the same observing run.  
As discussed in \S 3 below, the photometric uncertainty achieved in the survey 
is generally limited by the systematics of object detection, aperture corrections,
blending, and sky evaluation and not by the precision of the calibrations.  
We assumed atmospheric attenuation 
of 0.56, 0.19, 0.08, and 0.06 magnitudes per airmass for $U_n$, $G$, ${\cal R}$, and $I$,
respectively; these values were used to reduce the standards observed
at a variety of airmasses, and no significant departures from these values were
indicated (the internal scatter of the photometry is as good as it can be given
the precision of the spectrophotometry of the standards we have used). 
Effective zero points were calculated for each image by using the single
(photometric) exposure observed at smallest airmass in each band, and forcing the 
magnitudes in the stacked image to yield the same magnitudes as in the calibration image. 
This procedure mitigated uncertainties in the atmospheric attenuation constants and
allowed for the inclusion of non-photometric data in the stacks. The photometric zero
points were then corrected for Galactic extinction using the relations
$$A(U_n)=4.8E(B-V); \quad A(G)=3.7E(B-V); \quad A({\cal R})=2.3E(B-V);  \quad A(I)=1.7E(B-V)$$
where $E(B-V)$ was estimated from the relation $E(B-V)=0.018 S_{100}$ and
$S_{100}$ is the IRAS 100 $\mu$m intensity in MJy sr$^{-1}$. This extinction calibration is
very similar to that advocated by Schlegel, Finkbeiner, \& Davis (1998). The adopted 
Galactic extinction for each field is summarized in Table 2.    

The reduced images were astrometrically calibrated, in the case of data obtained at Palomar,
using the established plate solution for COSMIC and zero-pointed using several reference
stars on the Hubble Space Telescope guide star catalog system. All other data were astrometrically
calibrated with reference to the USNO-A2 astrometric catalog (Monet \et 1996), matching typically 50-100 objects
and using a polynomial solution to map the focal plane to the astrometric reference. The typical
residuals with respect to the USNO-A2 positions were 0.3-0.4 \arcsec, but we estimate that the internal
accuracy for the relative astrometry in all cases is better than $\sim 0.1$\arcsec\, based on
experience in the fine-alignment of slit masks.   

\section{GALAXY PHOTOMETRY}

We adopted a uniform method for performing the galaxy photometry which
was maintained throughout the survey; it is nearly identical to that
described previously in Steidel \& Hamilton (1993), Steidel, Pettini, \& Hamilton (1995),
and Steidel \et (1999).        

For all survey fields, the image quality was matched in each band by smoothing the
data with a Gaussian kernel so that the FWHM of stars was the same to within $\sim 5$\%.
In practice, this usually meant that all of the images were smoothed to match the image
quality in the $U_n$ band data. We used a modified version of the FOCAS (Valdes 1982)  photometry
program for the detection and photometry of the galaxies. Briefly, the program was
modified to allow for improved measurement of the local
sky values and to allow for masking of regions of the image in the detection process.

The detection of objects was performed on the ${\cal R}$ band
image, after lightly smoothing with the default 2-pixel FWHM smoothing kernel. 
Objects were ``detected'' if the number of connected pixels with flux exceeding 3 times
the sky $\sigma$ (of the unsmoothed data)  was such that their isophotal area
exceeded the size of the seeing disk (typically about 1 square \arcsec\ ). In practice,
even at the faint limit of our final catalogs (${\cal R}=25.5$ for all but Q1422+2309) the typical
isophotal size of a detected object is $\sim 3$ square arcseconds, so that the isophotal
aperture is roughly equivalent to a 2\arcs\ diameter circular aperture. The FOCAS
``total'' magnitude is formed by growing the isophotal detection aperture by a factor
of two in area, i.e. equivalent to a $\sim 3$\arcsec\ diameter aperture at the faint limit
of our catalogs.  The average aperture corrections (i.e., $m_{\rm iso}-m_{\rm tot}$) 
are nearly independent of apparent magnitude over the range used in the survey, and range
from 0.10-0.15 magnitudes for a typical LBG survey field. Simulations described below
show that using apertures of this size leads to negligible light loss.
The local sky background is evaluated in an ``annulus'' of specified width
(in pixels) outside of this total magnitude aperture, excluding any pixels that are within
the detection aperture of any other detected object. An attempt is made to de-blend objects
whose detection isophotes are merged by raising the isophotal level until separate objects,
each of which satisfied the minimum isophotal size criterion, are found. The light within
the parent total aperture is then divided among the child objects according to the ratio of the
luminosity within the raised isophotal apertures. The set of isophotal detection apertures
is then transferred to the images in the other passbands, so that colors are measured through
identical apertures.  When we quote a ${\cal R}$ magnitude, we refer to the FOCAS
``total'' magnitude, while the galaxy colors are measured using the isophotal
apertures. Thus, we implicitly assume that there are no significant color gradients in the
galaxies, which seems reasonable given that most of the objects faint enough to be LBG candidates
are barely resolved in $\sim 1$\arcsec\ seeing. A final culling of the photometric catalogs is made by requiring
that $19.0 \le {\cal R}\le 25.5$ (except for the Q1422+2309 field, which has a faint limit
of ${\cal R}=26.0$; the bright limit ensures that objects which might have been saturated
in individual exposures are not included), and that objects are detected in the $G$ band with greater than 3$\sigma$ 
significance (otherwise measured $U_n-G$ colors would have no statistical significance). No requirement is made on the $U_n$ magnitude. 

For some purposes, we have made use of the image statistics to evaluate the photometric depth
of the survey images. We define 1$\sigma$ limits 
by $N_{\rm pix}^{0.5} \sigma_{\rm sky}$, where $N_{\rm pix}$ is the number of pixels in the
detection aperture and
$\sigma_{\rm sky}$ is the RMS pixel to pixel fluctuations in the sky background\footnote{The RMS
values used for this calculation were corrected for pixel-to-pixel correlations 
introduced in the process of registering and re-sampling the data. Although high order interpolation schemes were 
used for the re-sampling, the actual measured values of $\sigma_{\rm sky}$ are too small
by $\sim 10$\%.}.  
Limits defined in this way are used to distinguish between objects that are ``detected'' in
the $U_n$ band and those that are merely limits. In our definition, if the $U_n$ flux in the
object aperture exceeds $1\sigma$ as described above, the object is considered ``detected'' and the
$U_n$ magnitude assigned is the value appropriate for the measured flux; if the flux in the aperture
is less than $1 \sigma$ the object is assigned a limit that is defined as the magnitude corresponding
to the aperture 1$\sigma$ limit. 
In Table 2 we show the 1$\sigma$ limits in each band for each field, for an aperture
having an area three times the size of the seeing disk;
this aperture approximates the typical
isophotal detection aperture for objects at the faint limit of our catalogs. 
In general, the faintest objects retained in the catalogs
(${\cal R}=25.5$, and ${\cal R}=26.0$ for the Q1422$+$2309 field) are $\sim 10\sigma$ detections in the ${\cal R}$ band.    
Thus, formally, we expect that the photometric errors for the ${\cal R}$ magnitudes are $\simlt 0.1$ mags
(but see below). The expected color 
uncertainties will of course depend upon the galaxy colors; for
objects with the typical colors of LBGs ($G-{\cal R} \simeq 0.5$, $U_n-G > 2.0$), the formal 
uncertainties should be smaller than 
$\sim 0.15$ magnitudes in $G-{\cal R}$ but (since most $U_n$ magnitudes are close to the 1$\sigma$ limits)
$\sim 0.6-0.7$ magnitudes in $U_n-G$ for objects at ${\cal R}=25.5$. 
At the catalog limit of ${\cal R}=25.5$, objects having the relatively blue $U_n-G$
colors of typical faint galaxies (see Table 2)
would have formal color uncertainties in $U_n-G$ of $\sim 0.25$ mags. 

Perhaps more realistic photometric uncertainties have been obtained from 
extensive simulations in which objects of known colors and magnitudes were added to the data 
and then recovered and measured using the same photometric procedures as 
for the real data. The results from such simulations have already been used in Steidel \et (1999), Adelberger \& Steidel (2000),
and Shapley \et (2001); they are described in detail in Adelberger (2002). 
The simulations show that measurement errors tend to be dominated by the
systematics of the photometry-- detection, blending, etc.--
and not by Poisson counting statistics of the fluctuations in the sky background as is often
assumed for faint galaxy photometry (see Steidel \& Hamilton 1993). 
The photometric uncertainties for detected objects
depend in a complex way on local environment, seeing, depth, color, and magnitude. 
In Table 3, we summarize the typical photometric uncertainties associated with objects having
the colors of LBG candidates (see discussion below). These were estimated by adding 
a large number of objects with colors and magnitudes similar to the LBGs of interest,
and examining the photometric statistics of the objects that were recovered 
as LBG candidates. Note
that the ``input'' colors and magnitudes need not have satisfied the 
LBG color selection criteria, and a considerable fraction of the objects with
input ``LBG'' characteristics are not recovered, for a variety of reasons.
Evaluating the photometric {\it completeness} as a function apparent magnitude
and color is a separate, important question, particularly for using the photometrically-selected
samples for inferring properties
of the ``parent population''-- see Steidel \et (1999) and Adelberger (2002) for more complete
discussions of this topic and applications of the technique. 

The data in Table 3 are an average over the fields included in the survey, and so
indicate typical photometric uncertainties.
Full modeling of these effects in each field is important for many
applications of the data; here we simply make a few general points.
First, it is often the case that the uncertainties in, e.g., $G-{\cal R}$ colors, are smaller than the errors in the
${\cal R}$ magnitudes (which are almost independent of magnitude). This is because, to first order,
the systematics that dominate the magnitude errors (crowding, local sky evaluation, blending)
are strongly correlated in the $G$ and ${\cal R}$ passbands. Since we are using identical apertures (defined in the
${\cal R}$ band) in each band, we suffer most from systematics in the detection phase, and then to a lesser
extent in measuring colors. In addition to increased uncertainties due to these systematics, there are
also {\it biases} that are introduced. Table 3 shows that, at the faint end of the LBG magnitude distribution, especially for objects with redder $G-{\cal R}$ colors, 
there is a tendency to measure galaxies to be systematically too red
in $G-{\cal R}$ by up to $\sim 0.1$  magnitudes. This is related to the fact that the
apertures are defined in the ${\cal R}$ band, and adjust themselves to positive fluctuations 
in the sky background near the sky level. Since the same aperture is applied to the $G$ band light,
the aperture is not free to adjust itself similarly to maximize the $G$ band light (see a more
extensive discussion in Steidel \& Hamilton 1993).  
Similar reasoning also explains the tendency to {\it over-estimate} slightly the
${\cal R}$ band flux for the brightest objects, although this may also be due to the fact that
larger isophotes are more likely to include light from neighboring objects that are not individually
detected.  

An estimate of the consistency of the photometry from field to field can be made by comparing
the colors and number counts of faint objects detected in each. For this particular benchmark,
we take all objects in the apparent magnitude range $22.5 \le {\cal R} \le 25.0$ in each field.
This range was chosen to be sufficiently faint so as to minimize fluctuations in the
foreground (low redshift) galaxy population and different stellar contamination
as a function of Galactic latitude, but bright enough that incompleteness at the faint end (that is
very seeing and depth dependent) is unlikely to be important.  The statistics on the surface density
and the mean $G-{\cal R}$ and $U_n-G$ colors (for objects significantly detected in all 3 bands) are
summarized in Table 2. The surface density
of objects in the ${\cal R}=22.5-25.0$ range has an average of $\simeq 26$ arc min$^{-2}$ and a field--to--field
scatter in the mean of $\sim 7$\% (see Table 2). Not surprisingly, there is a tendency for fields with better seeing
to have slightly higher surface densities, likely indicating differences in photometric completeness
at the $\sim 10$\% level (due to both increased depth and decreased blending) depending on the seeing. The slope of the ${\cal R}$ band number counts (Steidel \& Hamilton 1993)
is such that a zeropoint error of $0.1$ magnitudes would translate into a change in the
surface density of 7\%, so that at least some of the field-to-field scatter could be attributed
to slightly different ${\cal R}$ band photometric zero points. It is easy to imagine that uncertainties
in the Galactic extinction could modulate the zero points by up to $0.05$ magnitudes, particularly in fields
with significant Galactic 100 $\mu$m cirrus emission.   

The field-to-field scatter in the mean colors of the faint but well-detected objects used in the benchmark test
is $0.04$ magnitudes for $G-{\cal R}$ and $0.06$ magnitudes for $U_n-G$. In most cases, the photometric
calibration of each field is independent of the other fields (because it takes $\sim 2$ good nights to
image a single field, generally only one field was observed per typical observing run) so that much of this
scatter can be attributed to the uncertainty in the calibration of the photometric zero points. 
Additional scatter might result from uncertainties in the dependence of atmospheric absorption on airmass, differences
in the spectral response of the telescope plus detector system\footnote{We have seen clear evidence for large variations
in the UV reflectivity of telescope mirrors depending on time since the last aluminization, or when the
mirror has been subjected to accidental moisture condensation, etc.}, systematic differences in the calibrations of
different spectrophotometric standard stars, and uncertainties in Galactic extinction (we note that
$E(G-{\cal R}) = 1.4 E(B-V)$ and $E(U_n-G) = 1.1E(B-V)$). Finally, of course, there can be a contribution
to the field-to-field scatter in all of the quantities examined due to sample variance (i.e., ``cosmic scatter'')
that is difficult to quantify. In view of the many possible ways that the photometry might have been affected,
one might conclude that it is actually surprisingly consistent from field to  field. In any case,
no measures have been taken to adjust the photometry in any field excepting SSA22b, whose $G$-band zero
point  was adjusted
slightly to bring it into better agreement with its neighbor field, SSA22a (which had much more trustworthy calibration data).  
It may be important to keep in mind, for some applications, that systematic effects in both colors and magnitudes
at the $\sim 0.1$ mag level are possibly present in the photometric catalogs.

\section{PHOTOMETRIC SELECTION OF CANDIDATES}

The selection of objects in color-color space for spectroscopic follow-up has been dictated by
practical considerations and partially by expectations from simple models (which have been subsequently verified
empirically). Our filter system was originally intended for the efficient selection of objects
near $z \sim 3$ through its sensitivity to the Lyman continuum decrement passing through the 
near UV passband. The original criteria we used for flagging candidates (Steidel \& Hamilton 1992, 1993;
Steidel, Pettini, \& Hamilton 1995) were designed to be conservative in the sense that the part of
the $U_nG{\cal R}$ color space selected was deliberately steered well away from the stellar locus, 
and from the bulk of the so-called ``faint blue galaxies'' (e.g., Guhathakurta \et 1990). 
Inspection of a simple diagram such as the one in Figure 2, which shows the expected colors
versus redshift of model star-forming galaxies with a variety of assumed
reddening by dust, makes it clear that a selected region
is going to cut across any realistic galaxy distribution in a complex manner and that, in particular,
the region of color space where the galaxy $U_n-G$ colors are increasing rapidly is likely to also
mark galaxies at redshifts $z \simgt 2$. As discussed in Steidel \et (1999), and to a greater extent
in Adelberger (2002), color criteria will select a different subset of the star forming galaxy
population as a function of redshift, but careful attention to completeness issues as a function
of color, magnitude, and  redshift can allow one to reconstruct the true distribution of
the underlying ``population'' over the full range of intrinsic properties represented in the observed sample. 
The object of color selection is to select the maximum number of objects having the properties of
interest with the minimal contamination from objects at other redshifts.  

All the objects in our survey were chosen to satisfy the color criteria
$G-{\cal R}\le 1.2$, $(U_n-G) \ge (G-{\cal R}) + 1.0$, but for some purposes
(mainly historical)
we subdivided this region of color-color space into 2 regions, and in each we 
distinguish between
objects which are and are not detected in the $U_n$ band at the 1$\sigma$ level. 
A summary of the selection criteria we have used for the survey discussed in this paper (results from surveys
using other photometric selection criteria will be presented elsewhere) is given in Table 4, and these
regions are illustrated on the two-color diagrams in Figures 2 and 3. 
The region occupied by the
``C'' and ``D'' candidates is identical to that proposed for ``robust'' candidates in Steidel, Pettini, \& Hamilton 1995;
initial spectroscopic success in expanding these criteria to smaller $U_n-G$ colors without introducing
significant contamination from low-redshift objects prompted us to include the additional ``strip''
defined by the ``M'' and ``MD'' criteria.  Including this smaller region, which in practice has a much
higher ``density'' of candidates than the C/D region of color space, brings the surface density of candidates
to an average of $\sim 1.8$ galaxies arcmin$^{-2}$-- well matched to the density required for making
good use of the available multi-object spectroscopic capabilities at Keck. Evidently, based on more recent
exploration of still bluer (in $U_n-G$) objects with the LRIS-B instrument, one could relax the criteria
further still without paying a significant penalty by including many low-redshift interlopers. However,
as we discuss further below, for our initial spectroscopic follow-up, the galaxies in the redshift
range $2.7 \simlt z \simlt 3.4$ provide an almost ideal match to the spectroscopic capabilities that
were available during the time that these survey data were obtained. Including lower redshift objects
would have led to significantly higher rates of spectroscopic incompleteness. 

One can see in figure 2 that some contamination by Galactic stars is expected in both the C/D
and M/MD regions of the selected color regions. Over the magnitude range $19 \le {\cal R} \le 25.5$,
the G and K stars will all be main sequence stars. We discuss the stellar contamination in \S 6 below.  

Because of photometric errors and real color differences among selected objects at a given
redshift, 
the act of drawing what are to some extent arbitrary lines in color space
does not impose a hard redshift cutoff for the sample. Instead, the color cuts should be thought of as providing only
crude control over the redshift selection function, while at the same time allowing for samples
with a fairly broad range of intrinsic properties at a given redshift. This is a desired property
if one is interested in the statistics of star forming galaxies in general and not those with
a particular rest-frame properties. Since at the time we began this survey the $z > 2$
universe was largely unexplored, it seemed most prudent to simply ``see what is there'' rather
than fine-tune the selection criteria based on (perhaps unfounded) expectations. 

Still, small differences in photometry can have a significant effect on which objects
fall into the selection region, particularly near the ``boundaries'' of the windows:
for example, 0.1 mag differences can easily turn C/D objects into M/MD objects,
and the same differences can cause significant numbers of objects which belong in the M/MD
window to fall out of the observed sample. There are some fields in which our
photometry evolved over time,  either because deeper data were obtained or because significantly
improved calibrations became available. For this reason, in some fields we have
obtained spectra of a number of galaxies that were candidates in a previous 
incarnation of the photometry, but no longer are included in the ``official'' selection
windows. These objects, when redshifts were successfully measured, will be included for
completeness' sake in the LBG catalogs below, but are designated as ``old'' (e.g., HDF-oMD49) and
have not been used in any statistical calculations in this paper.

\section{SPECTROSCOPIC OBSERVATIONS AND DATA}

After the initial spectroscopic tests of the Lyman break technique (see, e.g., Steidel \et 1996a), 
we converged on a general approach that was followed for all subsequent spectroscopy. 
Having generated lists  of photometric candidates as discussed in the previous section, the candidates
were visually inspected to exclude obviously spurious objects (e.g., ``objects'' associated with asteroid
trails, diffraction spikes, bleed trails, etc.) and these objects were deleted from the lists. 
Each object was assigned a numerical priority, based primarily on apparent magnitude in the ``blank'' fields,
and on a combination of apparent magnitude and projected distance from the background QSOs in the
fields observed as part of the galaxy/IGM survey described in Adelberger \et (2003). Occasionally
objects of particular interest would receive especially high priorities to make sure they would
be assigned to a slit mask, but in general objects brighter than ${\cal R}=25$ were given roughly
two times higher priority than those fainter than ${\cal  R}=25$, and ``C'' and ``D'' type
candidates were given higher priorities than ``M'' and ``MD'' candidates. A comparison of the 
apparent ${\cal R}$ and $G$ magnitude distributions
of objects for which spectroscopy was attempted  
versus the full photometric sample is shown in Figures 4 and 5, respectively. Figure 6
shows similar distributions separated into the 
the C/D/M/MD subsamples. 

Slit masks were designed by allowing a computer program to maximize the total priority of
the objects that could be placed on a given slit mask. Usually sky position 
angle and mask center position were allowed to
vary in the optimization process. 
The program allowed slits to be placed anywhere within a field 7.3 by 5 arc minutes 
so long as they did not overlap in the spatial direction. The minimum acceptable slit
length was $\sim 9$\arcsec\, to allow for adequate sky subtraction for even the shortest slits. 
Once an object was selected for a mask, its priority was downgraded (or in some cases it was removed
completely) before designing the next mask for a given field. A typical slit mask included an average of 15--20 $z \sim 3$ LBG candidates. Often we would experiment with ``filler'' objects selected using different
color criteria to fill in any gaps in the slit mask ($\sim 10-15$ additional slits per mask). 
In this way we were able to explore how the use of
different color criteria would influence the resulting sample. In this paper, we report 
the results only for objects satisfying the color criteria described in \S 4. 
Our method for choosing objects for spectroscopy tended to favor sparse ($\sim 50$\%) sampling of the full
imaged region over complete spectroscopy in any given region. Examples of the spatial distribution of
objects targeted for spectroscopy versus the full distribution of candidates are shown in Figure 7.  

All but a few of the slit masks observed for the $z \sim 3$ LBG survey used the same instrumental configuration.
We used the Low Resolution Imaging Spectrometer (LRIS; Oke \et 1995) on both Keck I (October 1995-August 1996 and
January 2000-present) and Keck II (October 1996-November 1999). We used a 300 line mm$^{-1}$
grating blazed at 5000 \AA\ in first order, leading to a dispersion of 2.47 \AA\//pixel on
the Tektronix 2K$\times$2K CCD and a typical spectral coverage that was adjusted (with the grating
tilt) to include at least the
$4000-7000$ \AA\ range on every slit. The nominal spectral resolution in combination with the   
1\secpoint4 slitlets\footnote{These were considerably wider than the typical seeing
at the Keck Observatory, and were chosen to mitigate the effects of atmospheric dispersion and
errors in the relative astrometry. With hindsight they were probably wider than they needed to be.}
is $\sim 12.5$ \AA\ (FWHM) evaluated from the width of night sky emission lines. However, given that most
of the galaxies are barely resolved even in good seeing, the actual spectral resolution is ``seeing-limited''
in almost all cases. The typical image quality at the detector was $\sim 0.8-0.9$\arcsec\ (FWHM), so that
the effective spectral resolution (neglecting guiding errors) is closer to 7.5 \AA\ in most cases.
No attempt was made to observe masks with position angles close to the parallactic angle, and
we sometimes observed masks to airmasses of 1.5-1.6 when fields were setting. Thus, there is some danger
that the apparent object position perpendicular to the slit could be wavelength dependent particularly
when the seeing was very good.

A small subset of the data in the Q0933+288 and Q1422+2309 fields, and all of the data in Q0302-003, were obtained
using the new blue arm (LRIS-B; McCarthy et al 1998, Steidel \et 2003, in preparation) of LRIS between late 2000 and early 2002. 
Most of these observations 
were obtained by directing all of
the incident light into the blue channel using a mirror in place of the usual dichroic beam-splitter. The
disperser was a 300 lines mm$^{-1}$ grism blazed at 5000 \AA\, providing a typical wavelength coverage
of 3500-7500 \AA\ for each slitlet. The spectral resolution is slightly higher than that of the
LRIS-R data, while the spectral throughput is significantly higher. During the time these LRIS-B
data were obtained, the detector was an engineering grade Tektronix 2k x 2k UV/AR coated CCD.  

The observations for each slit mask were obtained in a series of 1800s exposures. The telescope
was dithered slightly ($\sim 1-2$\arcsec ) in the slit direction between exposures in order
to sample different parts of the detector and to allow for a variety of options for subtracting the
sky background. Initially, we obtained total integration times of $2-3$ hours per mask, but we found
that under good observing conditions (seeing FWHM$<0.8$\arcsec, clear) we could easily obtain
adequate data in a total integration time of 5400s. Flat fields were obtained at the end
of the observing sequence of a given mask using a halogen lamp internal to the spectrograph.
An observation of internal arc lamps (Hg, Ne, Ar, Kr, Xe) was also obtained at the end of
the observing sequence for a given mask.  

The data were reduced using a custom package based on IRAF scripts. Each observation of a given
slit mask was ``cut up'' into individual spectrograms and flat-fielded using the appropriate (continuum-normalized)
image section of the flat-field image. Cosmic ray identification was done using a routine
that identifies cosmic ray hits morphologically and adds them to a mask that is maintained for each
exposure and for each slitlet's spectrogram. The background was subtracted by fitting a polynomial
at each dispersion point whose order was varied depending on the length of the slitlet (the slitlets
varied in length from $\sim 9$\arcsec\ to $\sim 20$\arcsec).  The two-dimensional,
background-subtracted spectrograms were then shifted into registration using the positions of
night sky emission lines in the spectral direction (to remove the effects of any flexure during the 1.5-2.5 hours a given
field was being observed) and object continuum positions in the spatial direction, and averaged. Pixels
that had been masked as cosmic ray hits were excluded in forming the final stacked spectrograms.
The arcs corresponding to each slitlet were reduced in exactly the same way as the data (excepting the
background subtraction). The final one-dimensional spectra were then traced and extracted from the
background-subtracted final stacks; one-dimensional arc spectra were extracted from the same regions. The wavelength
solutions were then obtained by fitting a 5th-order polynomial to the arc lamp spectra; typical
residuals were $\sim 0.2$\AA. Finally, each 1-D spectrum was extinction corrected and (approximately)
flux calibrated using observations of spectrophotometric standards observed with a 1\secpoint5
long slit and the same 300/5000 grating used for the slit masks, and then reduced to vacuum 
wavelengths to facilitate redshift measurement using far-UV transitions.  

\subsection{Spectral Identification}

As has been discussed extensively elsewhere (e.g., Steidel \et 1996a, 
Steidel \et 1998, Pettini \et 2000, 2001), 
most of the easily identifiable features in LBG spectra are interstellar absorption lines
of very strong transitions in the $1200-1700$\AA\ range in the rest frame. 
The dark sky background (and high instrumental efficiency) in the $4000-6500$ \AA\ 
range and the typical 5-10 \AA\ equivalent
widths of a number of easily recognizable absorption features makes it surprisingly easy to
measure redshifts even from spectra having a signal--to--noise ratio ($S/N$) per resolution element of only a few. 
The process is also greatly assisted by the nature of these absorption features. As described in
detail in Shapley \et 2003, objects with the strongest Lyman $\alpha$ emission lines tend to have
weaker interstellar absorption features (and bluer continua) while objects having weak Lyman $\alpha$ in emission or
Lyman $\alpha$ completely in absorption tend to have significantly stronger interstellar absorption
lines.  

The spectra for the LBG sample span a large range in quality, from objects with single
emission line detections to those with many identified absorption lines. Here we outline
the criteria for assigning redshifts from the spectra. 
Every observed spectrum was examined interactively (in both 1-D and 2-D) by at least
two of us; the same two people examined the whole sample several times each. 
Approximate redshifts were first assigned by identifying a single feature and marking the
expected positions of other strong far-UV features.   
More precise redshifts were then measured; in the case of Lyman $\alpha$ emission lines by
fitting a Gaussian to the observed profile, while for absorption features the average redshift
given by the centroid positions of all well-detected features was adopted.
If a single emission line was observed,
especially in combination with a discernible continuum drop shortward of the line, the
redshift was considered secure. In general, a single emission line is insufficient to
reliably identify spectra, but in combination with the photometric break that placed the
object into the sample, a single emission line is extremely unlikely to be anything except
Lyman $\alpha$, particularly if that line suggests a redshift consistent with the
continuum break. 

The spectra generally fall into 3 fairly distinct classes: those
that are identified using Lyman $\alpha$ emission and broad continuum properties alone,
those that are identified by means of multiple interstellar absorption lines, and those
that have both Ly $\alpha$ emission and one or more clearly identified absorption lines.
Examples of each type are shown in Figure 8; these spectra were chosen also to be representative
of the range of quality present in the full sample. The precision with which the redshifts are
measured, based upon independent measurements of the same spectra or on the scatter
among different lines, is 
$\Delta z \simeq 0.002$ for absorption line measurements and $\Delta z \simeq 0.001$
for emission line objects. The larger dispersion for the absorption line objects reflects
the fact that these lines are often quite broad compared to the emission lines and have lower
S/N.  As has been discussed extensively elsewhere 
(e.g., Franx \et 1997; Steidel \et 1998; Pettini \et 1998, 2000, 2001, 2002;
Adelberger \et 2003; Shapley \et 2003) the Ly $\alpha$ emission and interstellar absorption features
are almost universally separated by at least several hundred \kms. This phenomenon is generally
interpreted as evidence for strong outflows from LBGs, where the interstellar absorption lines
are produced by the near side of the outflow, and the Ly $\alpha$ emission line by the back-scattering
from the opposite side of the outflow, and
where the true systemic redshift of the galaxy lies somewhere in between. For this reason, we have
recorded both emission $(z_{\rm em})$ and absorption $(z_{\rm abs})$ redshifts for each
galaxy, when both have been measured. Methods for obtaining more precise systemic redshifts
from the far-UV spectra alone (and the resulting uncertainties) 
are discussed in Adelberger \et (2003).  

Table 5 summarizes the total number of objects observed in each of the 17 fields. The overall spectroscopic
success rate is $\sim 76$\% (including 3.5\% spectroscopically identified interlopers); most of the 24\% 
of failed measurements were obtained under
less than ideal conditions, whereas masks obtained under good or very good observing conditions
(clear skies, seeing FWHM $\le 0.8$\arcs)  
tended to reach greater than 90\% success, where success is defined as the fraction of attempted
objects that have yielded redshifts. Note that in figure 4, the spectroscopic success rate
actually reaches a minimum in the ${\cal R}=24.5-25$ range, and then {\it increases} toward
fainter magnitudes beyond ${\cal R}\sim 25$. While somewhat counter-intuitive, it is due to
the nature of our photometric selection criteria. For the faintest objects in the sample, the
demands of minimum ``spectral curvature'' criteria (i.e., the difference between the $U_n-G$ and
$G-{\cal R}$ color) impose significant dynamic range constraints at the faint end of the apparent
magnitude range considered. Only the bluest objects at ${\cal R}=25.5$ can be selected using
our color criteria because of the finite depth of the $U_n$ band images. The bluest LBGs almost
invariably have Lyman $\alpha$ in emission, so that redshifts are usually measurable even when 
there is little or no (spectroscopically) detected continuum. The relationship
between color and spectral properties is discussed in much more detail in Shapley \et (2003).
Figure 9 shows composite spectra
formed from each quartile of the spectroscopic sample divided by the equivalent width of the
Lyman $\alpha$ feature. A detailed discussion of the astrophysics that can be extracted from
these and other LBG composite spectra is contained in Shapley \et 2003. 

In some cases, the quality of the observed spectrum is inadequate to assign what
we consider to be a precise redshift, but is good enough for a reasonably secure one. 
Usually these spectra are of high enough quality to note the position of a ``break''
in the continuum that lies near the wavelength of Lyman $\alpha$ (caused by the blanketing of
the Lyman $\alpha$ forest) and where 1 or more other (low S/N) plausible 
spectral features can be identified, but where none of the additional features is secure enough
to provide a very high degree of confidence. Experience with repeat observations of insecure
redshifts suggests that about 80\% of them are very accurate, with the remainder being incorrect
by as much as $\Delta z \sim 0.1$. We have attempted to use cross-correlation techniques to improve
on the redshift identifications of redshift failures or insecure redshifts but have found
that the results lead to an unacceptable incidence of 
spurious identifications; thus, we have decided to err on the side of caution. 
Objects with less certain redshifts have been flagged in the catalog tables 
(with redshift entries preceded by
a colon, e.g. ``:2.789" in tables 7--23), and are not
used in any analyses that depend on precise redshift measurements. Of the total
of 955 objects satisfying the LBG photometric criteria with spectroscopic redshifts $z > 2$,
121 fall into this less precise category. There is no significant difference in the overall
redshift distribution of the 834 class 1 redshifts compared to the 121 in class 2.  

A redshift histogram for the full spectroscopic sample is shown in figure 10. Also shown are the
separate histograms for the 4 ``types'' of LBG candidates, to illustrate overall redshift
differences depending on the location in the color-color plane of the candidates.   
The redshift statistics are summarized in table 6. 
The general trends with color can be understood as follows (see figure 2 for a graphical depiction):
objects of a given intrinsic spectral energy distribution (SED) will become redder in both $G-{\cal R}$
and $U_n-G$ with increasing redshift over the range spanned by the survey. The $G-{\cal R}$ effect is due to increased blanketing
of the $G$ passband by the Lyman $\alpha$ forest, and the $U_n-G$ effect is caused by an increasingly large
fraction of the $U_n$ passband falling shortward of the rest frame Lyman limit (for $z > 3.3$ the $U_n$
passband is entirely shortward of this rest wavelength). Because of the finite depth of the $U_n$ images, 
we normally cannot measure a $U_n-{\cal R}$ color greater than about 3 magnitudes at the faint end of
the ${\cal R}$ magnitude distribution. The limits on $U_n-G$ color will then clearly depend on the 
measured $G-{\cal R}$ color. 
The slightly higher redshifts of ``C'' type 
candidates compared to ``D'' objects, which differ only in whether they are detected at better
than the 1$\sigma$ level in $U_n$ (see \S 3 and table 2), are best understood as being due to 
the very strong redshift dependence of the $U_n-G$ color. 
As expected (see Figure 2), the ``MD'' type objects lie at somewhat lower redshifts
on average than the ``C'' and ``D'' type candidates-- the $U_n$ band
is less absorbed by the IGM and the Lyman limit at smaller redshifts.
The ``M'' type candidates, which are objects with less stringent constraints on the spectral
curvature (measured by the difference between $U_n-G$ and $G-{\cal R}$ color) are at {\it higher} redshift
than even the ``C'' candidates. This is most likely due to the dynamic range problem mentioned above; ``M''
candidates tend to be objects that are too red in $G-{\cal R}$ to result in stringent limits on $U_n-G$,
evidently driven mostly by the fact that they lie at higher redshifts where the red color is due primarily 
to forest blanketing. 
The LBG selection function, formed by the sum of all of the
candidate types, reflects the overall sensitivity of the survey to galaxies as a function of
redshift.  We emphasize that the shape of the LBG selection function is to some extent
dictated by the sampling rate of the $U_nG{\cal R}$ color plane which, as we have discussed,
favored (by fraction) objects with larger limits on or measurements of the $U_n-G$ color. 
Table 6 summarizes the sampling and success rates for the various candidate types; from this,
it can be seen that C/D candidates enjoyed a selection rate approximately 50\% higher than
M/MD candidates, and had a spectroscopic success rate that was slightly higher as well. 
Proper accounting of both the photometric biases and the spectroscopic sampling rate is necessary
for many applications of the LBG sample (see, Adelberger \& Steidel 2000, Steidel \et 1999, Shapley \et 2003).


\subsection{Field Redshift Distributions}

Figure 11 shows redshift histograms for each of the 14 distinct sky regions surveyed.
Here we have combined fields with two adjacent pointings (DSF2237a,b; SSA22a,b; CDFa,b)
into single histograms. In each panel, the light colored histogram shows the expected
redshift distribution given the overall survey selection function; thus, the figure
illustrates qualitatively the clustering properties within individual fields and the variance
of the large-scale redshift distribution from field to field. Because some of the
fields were selected to surround known QSOs or AGN (Q0302$-$003, Q0201$+$1120, Q0256$-$000,
Q0933$+$289, B20902$+$34, Q1422$+$2309, and Q2233$+$1341), we have marked the redshift
of the known objects in each case.  

Table 5 summarizes the spectroscopic results in each of the 17 fields, including
the fraction of candidates that were spectroscopically observed, the fraction yielding
redshifts $z >2$, and the fraction of ``interlopers'' or contaminants.  

\section{CONTAMINATION OF THE LBG SAMPLE}

A total of 40 stars are spectroscopically identified in the full LBG sample, or 
$\sim 4$\% of the total spectroscopic sample. The colors and spectra of these stars (see figures 2, 3) suggest
that most are Galactic K sub-dwarfs (i.e., halo main sequence stars). 
In figure 12, we plot the histogram of observed stars as a function
of apparent ${\cal R}$ magnitude. 
As can be seen from the figure, the stellar number counts are essentially
flat from ${\cal R}=22$ to ${\cal R}=24$, but then exhibit an apparent cut-off fainter
than ${\cal R}\sim 24.0-24.5$. There is essentially no identified stellar 
contamination of the LBG sample fainter than ${\cal R}=24.5$; however, we caution that 
the spectra of very faint stars may be more difficult to identify than LBGs of the same
apparent ${\cal R}$ magnitude, since there are generally fewer strong features in the
stellar spectra. Stars are also somewhat {\it under-represented} for objects brighter
than ${\cal R} \sim 23$, since when objects were obviously stellar {\it and} 
had colors on the stellar locus, they were given lower weights than other objects of
the same apparent ${\cal R}$ magnitude (this effect can be seen in Figure 6
as a smaller fraction of spectroscopically observed objects in the brightest bins   
for ``MD'' type candidates; the ``MD'' region of color-color space contains most
of the stellar interlopers, as shown in Figure 3). 



Of the 955 objects with redshifts $z >2$ and ${\cal R}\le 25.5$ in the sample satisfying the LBG color criteria 
28 (3\%) have obvious signatures of AGN in their
spectra. The AGN sub--sample is certainly interesting in its own right, and
is discussed in more detail in Steidel \et (2002); here
we discuss it only as a source of contamination of the LBG sample.  
All AGN or QSOs that were known prior to the survey and which were deliberately placed
within the field of view have been excluded from any numbers quoted below (i.e., the primary
QSOs in the QSO fields, and the radio galaxy B20902+34, are excluded). 

Of the 9 objects with $z > 2$ and ${\cal R}\le23$, 7 are identified as either broad or narrow-lined
AGN, while AGN comprise 8 of 31 high redshift objects (i.e., non-stars)
brighter than ${\cal R}=23.5$. Thus, as for the stars, 
the AGN contamination fraction is highly magnitude dependent. AGN have
generally been excluded from any published statistics of LBGs (e.g., clustering, luminosity
functions, etc.), although we have shown in Steidel \et (2002) that the AGN in the sample
are plausibly hosted by objects similar to Lyman break galaxies. 

Aside from the 40 stars,
there are only 5 other identified objects in the spectroscopic sample having $z < 2$, two
of which have $z\simeq 1.99$. The other 3 ``interlopers'' all have $z \sim 0.5$, and all
three are in the ``less secure'' class of redshift. The very low $z \le 2$ interloper fraction can
be attributed to a relatively conservative color selection window, and to some extent may be
caused by the limited spectral coverage of the typical survey spectrum, which would generally have
had difficulty identifying galaxies having $0.9 \simgt z \simlt 2.1$. However, we believe that
most of the identification failures have redshifts consistent with the LBG selection function but
failed because of inadequate S/N, in the majority of cases due to relatively poor observing conditions. 
As discussed above, the primary basis for this belief is that the masks observed under the
best conditions often achieved better than 90\% spectroscopic success rate, whereas there
were many poor masks on which only a few of the brightest galaxies were identified;
it was often impossible to re-observe objects that happened to be assigned to poorly-observed masks. 

\section{NOTES ON INDIVIDUAL FIELDS}

In the interest of completeness, we have included results for all of the survey fields observed during the course
of the $z \sim 3$ Lyman break survey. While the intended uses 
of the survey fields has varied, the same selection criteria and general survey approach
were used for all 17 fields. Several of the survey fields are 
``blank'' fields chosen either because other surveys had been or will be conducted there (e.g.,
Westphal, HDF-N, CDFa)
or as specially selected fields at particular RA that would be accessible
during scheduled observing runs with minimal Galactic extinction and very bright stars (e.g., 
DSF2237a and DSF2237b). Several other fields are pointings adjacent to initial
survey fields, chosen to increase the angular extent beyond the 9\arcm\ fields
afforded by the Palomar prime focus imager COSMIC (SSA22b, DSF2237b, CDFb). In all cases we have treated
these additional, adjacent fields independently, since both the imaging and
spectroscopic data were generally obtained on different observing runs and as a result the data
had somewhat different seeing, exposure times, and depth depending on the observing conditions and
available observing time. 
Five of the fields were centered on background QSOs
suitable for high resolution spectroscopy to be used in a comparison of the galaxy
distribution with the IGM along the same line of sight (Q0256$-$000, Q0302$-$003, 
Q0933$+$289, Q1422$+$2309, Q2233$+$1341), some of the results of which are presented
elsewhere (Adelberger \et 2003).  Three of the fields (Q0000$-$263, 3C324, Q0201$+$1120)
are included here but have generally not been used as part of statistical studies
either because of their small size (in the case of 3C324 and Q0000$-$263) or
because of excessive Galactic extinction (in the case of Q0201$+$1120).  
More details on each field are given below; see also Tables 1, 2, and 6.

Tables 7-23 contain the complete catalogs for all 17 of the survey fields. Entries of
$-1.000$ in the redshift column indicate that a candidate has never been attempted
spectroscopically; $-2.000$ in both the $z_{em}$ and $z_{abs}$ columns indicates that
the object has been observed spectroscopically but no reliable redshift resulted. 
Objects whose redshift measurement depended only on emission lines (usually Lyman $\alpha$ only,
for all but the AGN) have a $-2.000$ in the $z_{abs}$ column, and those objects without
measurable emission lines have $-2.000$ in the $z_{em}$ column. Redshift entries preceded
by colons indicate that the measurement is uncertain, as discussed in \S 5.1.  
The distinction between objects that are ``detected'' in the $U_n$ band and those that
were assigned a $+1\sigma$ limit is not made in the table entries; we refer the reader to 
Table 4 for a summary of the way in which the tabulated $U_n-G$ color should be interpreted, depending
on the candidate type. The significance of a particular $U_n-G$ color measurement can be judged
by comparing the $U_n$ magnitude [i.e., ${\cal R}+(G-{\cal R})+(U_n-G)$] to the tabulated values
of $\sigma(U_n)$ listed in Table 2 for each field.  Most of the $U_n-G$ values, being either limits
or close to limits, are uncertain by $0.4-0.6$ magnitudes, depending on color and apparent
${\cal R}$ magnitude.   

\subsection{Q0000$-$263}

The field of this $z_{em}=4.10$ QSO was included in pilot studies of LBG
search techniques (Steidel \& Hamilton 1992, 1993). These data were obtained
subsequently at the ESO NTT in order to improve significantly on the 
depth and seeing of the original data obtained at CTIO. The complete catalog of LBGs
in this field is given in Table 7. Some of the spectroscopic
results in this field were presented in Steidel \et 1996a--the
designation of the candidates has since changed, but cross-references to the old
names are given in the table. 
One note of caution is that the G and ${\cal R}$ filters used for the NTT data
are somewhat different in both center wavelength and bandpass from the filters
used for the rest of the fields; as a result, the photometry and candidate
selection is expected to have slightly different systematics compared to other
fields in the survey. The Q0000$-$263 field has generally not been used for more 
recent statistical studies of LBGs.

The object Q0000-C7 (see Table 7) is the emission line galaxy ``G2'' discovered via narrow-band
Lyman $\alpha$ imaging by Macchetto \et 1993 and further discussed by Giavalisco \et 1994, 1995.  

\subsection{CDFa,b}

CDFa is centered on the ``Caltech 0 Hour Redshift Survey Field'' discussed by Cohen \et 1996, which
itself is centered on a relatively deep HST/WFPC-2 pointing. CDFb is an adjacent field to
the South (and slightly to the East to avoid a bright star). The complete LBG catalogs for
these fields are presented in tables 8 and 9. 

\subsection{Q0201$+$1120}

The data obtained in this field were discussed by Ellison \et (2001), which 
presented results on the $z=3.390$ damped Lyman $\alpha$ system in the spectrum
of the $z=3.605$ QSO. It became clear to
us after the imaging data were obtained that the field suffers from quite heavy Galactic
extinction (amounting to $\sim 0.7$ mag in the $U_n$ band--see Table 2), although the photometry,
after nominal correction for extinction, appears to agree well with that in other fields. 
Nevertheless, we have not used Q0201$+$1120 for statistical studies of LBGs due to lingering
uncertainties about the quality of the photometry and the high probability of patchy (i.e., 
strongly variable) extinction over the field. Table 10 contains the LBG catalog for this field.

\subsection{Q0256$-$000}

The field is centered on the $G=18$ QSO with $z_{em}=3.364$, and was observed as part
of a survey to compare distribution of C~IV and H~I in the IGM to the galaxy distribution
in the surrounding volume (Adelberger \et 2003). The LBG data are summarized in Table 11.

\subsection{Q0302$-$003}

Another field observed as part of the galaxy/IGM survey, the QSO ($z_{em}=3.281$) is also one
of the few lines of sight that has been observed in the far-UV to probe
the re-ionization of He~II near $z \sim 3$ (Heap \et 2000). While a 15\arcm\ region was observed
photometrically, only a $\sim 7$\arcm\ region in the vicinity of the QSO has
been observed spectroscopically to date. The data in this paper include only
the region covered spectroscopically, although full photometric catalogs are
available on request. Table 12 provides the catalog over the spectroscopically observed region. 

\subsection{B20902+34}

This field is centered on the famous $z=3.392$ radio galaxy B20902+34 (Lilly 1988; the radio
galaxy is itself a LBG candidate, object
B20902-D6 in Table 13). It was observed to investigate the density of LBGs around
a high redshift radio galaxy, given the conventional wisdom that radio galaxies
inhabit rich environments. As can be seen from Figure 11, there is not
a highly significant galaxy over-density at the redshift of the radio galaxy, although
the statistics are not very constraining given the relatively small number of spectroscopic redshifts
in the field and the fact that the LBG selection function is rapidly declining 
at $z \sim 3.4$. The photometric data in this field are a combination of data obtained
at the William Herschel telescope prime focus imager and the P200+COSMIC at Palomar;
the field size is limited by the scale of the WHT imager at the time the data were
obtained.  

\subsection{Q0933$+$2854}

This field was selected because of its low galactic extinction and the presence of
the $z_{em}=3.428$, $G=17.5$ QSO, making it ideal for the galaxy/IGM survey project. 
The photometric data are a combination of KPNO 4m+Mosaic Camera data and data from
the Palomar 5m+COSMIC system; the region presented in this paper is the region common to
the two data sets, limited by the size of the COSMIC field. The LBG catalog is presented
in Table 14.   

\subsection{HDF-N}

We confine the data set for this paper to objects that were identified
only on the basis of the ground-based imaging from the Palomar 5m$+$COSMIC,
and not on other high redshift objects identified on the basis of the deep
WFPC-2 imaging near the center of the 8\minpoint7 field (cf. Steidel \et 1996b,
Lowenthal \et 1997; Dickinson 1998). Where there is overlap
between the HST-identified LBGs and the ground-based LBG survey we have provided
a cross-reference in Table 15. The imaging data in the HDF-N were obtained under
highly variable conditions, and hence are among the lowest-quality data in the
survey (e.g., the $U_n$ image is the shallowest of any of the 17 fields--see
Table 2).  

The X-ray properties of the (ground-based) LBGs in this field have been discussed by
Nandra \et (2002). 

\subsection{Westphal}

This field is named after James Westphal, the PI for the HST/WFPC-2 observation
of the deepest pointing of the
``Groth Strip'' WFPC-2 mosaic.  
The 15\arcm\ field includes the north-east section of the Groth strip mosaic,
and also several other relatively deep pointings of WFPC-2. 
This field contains the largest
number of spectroscopically confirmed LBGs in the survey (188), and the highest level
of spectroscopic completeness (relative to the photometric sample) of any of the
``blank'' fields. The field contains the entire Canada-France Redshift Survey (Lilly \et 1996)
14 hour field, and will be the subject of a number of current and future deep observations
at other wavelengths, including Chandra X-ray Observatory (200ks) and the Space
Infrared Telescope Facility (SIRTF). The LBG catalog is presented in Table 16. 

\subsection{Q1422$+$2309}

This field was chosen for the galaxy/IGM project, 
and is centered on the gravitationally-lensed $z_{em}=3.620$ QSO ($G=16.5)$. We intended 
to obtain especially
deep photometric data in this field because the spectra of the QSO
are exceptionally good (it is perhaps the most-observed high
redshift QSO in the sky). Pristine observing conditions at the
William Herschel Telescope coupled with an EEV/Marconi CCD that provided high UV quantum
efficiency allowed us to reach about 1 magnitude deeper in the $U_n$ band than in
any of our other fields. Hence, we were able to extend our selection criteria for LBGs
to ${\cal R}=26$ rather than our usual limit ${\cal R}=25.5$. In addition to the 453 LBG candidates
identified in this field, we also discovered a new $z_{em}=3.629$, ${\cal R}=22$
QSO only 40\arcs\ from Q1422$+$2309 itself. This object, dubbed Q1422$+$2309b, is
included in table 17 despite the fact that it does not quite satisfy
the LBG selection criteria (it is slightly too red in $G-{\cal R}$).   
 
\subsection{3C 324}

Only one slit mask was observed in this field, but it is included for completeness. 
It was originally observed to coincide with a very deep HST/WFPC-2 observation of
the radio galaxy field, but the positions of very bright stars forced moving the
pointing to the extent that the overlap with HST is now rather small.  The
LBG catalog is presented in table 18.

\subsection{SSA22a,b}

The SSA22a field was originally chosen to include several HST/WFPC-2 pointings
obtained as part of the Hawaii Deep Survey (e.g., Lilly, Cowie, \& Gardner 1991; 
Songaila \et 1994) and the Canada-France Redshift Survey (Lilly \et 1996). 
Some results from LBG observations in this
field were presented in Steidel \et 1996a; in addition, a prominent
redshift ``spike'' at $z=3.09$, interpreted as a proto-cluster region, was analyzed
by Steidel \et 1998, and followed up with very deep narrow band imaging, reported in
Steidel \et (2000). Sub-mm follow-up of the field is described in Chapman \et (2001).
The candidate designations in SSA22a have changed compared to their
original designations in these earlier papers; cross-references to the old names
are given in table 19. 

SSA22b is 8.5\arcm\ south of SSA22a, and was observed to increase the transverse scale
of the field to $\sim 9\times 18$ \arcm\. There are no HST pointings within SSA22b.   
The LBG catalog is presented in table 20.

\subsection{DSF2237a,b}

These fields were chosen because the best observing conditions at
Palomar generally occur in August and September. The fields were chosen to be in a region
of relatively low Galactic extinction, at high enough declination to be efficiently observed
from both Palomar and Mauna Kea, and without stars bright enough to cause scattered light
problems. There are no
ancillary observations of these fields using other instruments or at other wavelengths,
to our knowledge. DSF2237b is placed 8.7\arcm\ due west of DSF2237a,
to create a $\sim 9\times18$ \arcm\ field.  The LBG catalogs for the two fields are
presented in tables 21 and 22.

\subsection{Q2233+1341}

This field is centered on the $z=3.210$, ${\cal R}\sim 18.5$ QSO, and was observed as part of
the $z \sim 3$ galaxy/IGM survey. The LBG catalog is presented in table 23.   

\section{SUMMARY AND DISCUSSION}

We have presented the basic photometric and spectroscopic data for a large spectroscopic
survey of color-selected objects near $z \sim 3$. The sample was constructed using very
deep images in 3 passbands ($U_n$,$G$, and ${\cal R}$) in 17 different fields, covering a total
solid angle of 0.38 square degrees. The color selection criteria were designed to
isolate a well-defined redshift range for objects whose intrinsic spectral energy distributions
are relatively blue in the rest-frame wavelength range $1200 \simlt \lambda \simlt 1700$, so that
the marked discontinuity near the rest frame Lyman limit of hydrogen can be recognized using
broad-band photometry.  

Of the 2347 LBG candidates to ${\cal R}=25.5$ satisfying the color selection criteria, 55\% have been 
observed spectroscopically using LRIS on the Keck telescopes; of these, 76\% have
been spectroscopically identified and 73\% have measured $z > 2$. The redshift distribution
of the sample is approximately Gaussian, with 
$z=2.96 \pm 0.29$.  Approximately 4\% of the identified objects are Galactic stars, while
$\sim 3$\% have obvious spectral signatures of AGN (broad or narrow high ionization emission lines)
and have a redshift distribution consistent with that of the star-forming galaxies. 

We provide a full catalog of the 2347 LBG candidates (plus additional candidates to ${\cal R}=26$ in
one field), including coordinates, photometry and, for
those followed up spectroscopically, redshifts. Such data may be useful for future statistical
studies of this population of high redshift galaxies. 

This $z \sim 3$ survey for LBGs is one example of the type of approach to galaxy redshift
surveys that is necessary for efficiently studying the high redshift universe. We estimate
that the use of photometric pre-selection has improved the efficiency of studying galaxies
near $z \sim 3$ by approximately a factor of 30 as compared to a traditional apparent magnitude 
selected survey to the same apparent magnitude limit. With similar approaches, using new wide-field 
imaging
spectrographs coming on line on 8m-class telescopes, it would be within the bounds of reason
to undertake surveys of the high redshift universe as ambitious as state-of-the-art 
surveys of local galaxies (e.g., 2DF and Sloan Digitial
Sky Survey). It is also feasible to use suitably designed photometric
pre-selection for other redshift slices that will allow statistical samples of galaxies, isolating 
particular cosmic epochs, to be assembled rapidly. The obvious {\it caveat} is that these surveys
cannot possibly take in all galaxies present at a particular redshift, and so they cannot be treated
as all-encompassing; however, the same might be said of any survey. Insight into the process of galaxy
formation and the development of large scale structure comes from making optimum use of available
information and understanding how apparently disparate observations are related. The type of survey
described above offers a large amount of statistical information that is relatively easily
obtained using existing observational facilities, and may help guide future work as facilities
and understanding improve.

\bigskip
\bigskip
We thank Melinda Kellogg, Matthew Hunt, and Dawn Erb for help with some  
of the data presented.  
CCS, KLA, and AES have been supported by grants
AST-9596299 and AST-0070773 from the U.S. National Science Foundation and by the David and Lucile
Packard Foundation. KLA acknowledges support from the Harvard Society of Fellows.
We are especially grateful to the staffs of the Palomar Observatory, the Kitt Peak National
Observatory, the William Herschel Telescope, and the W.M. Keck Observatory for assistance with  
the observations. We benefited significantly from software developed by
Judy Cohen, Drew Phillips, Patrick Shopbell, and Todd Small. We thank the entire team
responsible for the Low Resolution Imaging Spectrometer, the instrument that has
made this work possible.  
We wish to extend special thanks to those of Hawaiian ancestry on
whose sacred mountain we are privileged to be guests. Without their generous
hospitality, many of the observations presented herein would not
have been possible.


\begin{deluxetable}{lccclclrcr}
\rotate
\tabletypesize{\scriptsize}
\tablewidth{0pc}
\tablecaption{Deep Imaging Observations}
\tablehead{
\colhead{Field Name} & \colhead{RA(J2000)\tablenotemark{a}} & \colhead{Dec(J2000)\tablenotemark{a}} & \colhead{Filter}  & \colhead{Telescope \& Date\tablenotemark{b}} & \colhead{Scale} & 
\colhead{FWHM} & \colhead{Exp Time} & \colhead{Dim} & \colhead{Area}
} 
\startdata
 Q0000-263 &   00:03:25.01 & $-$26:03:36.8 & $U_n$ & NTT 1994 Oct & 0.268 & 1.18  & 27600 & $3.69\times5.13$ & 18.9 \\
           &   & & $G$   & NTT 1994 Oct &       & 1.04  & 10800 &                  &      \\
           &   & & ${\cal R}$ & NTT 1994 Oct &  & 1.00  & 7200  &                  &      \\
           &   & & $I$   & LCO 1996 Jul  & 0.245 & 0.82 & 15600 &                  &      \\
 CDFa      &   00:53:22.97 & $+$12:33:46.3 &$U_n$ & P200 1996 Oct & 0.283 & 0.98 & 23400 & $8.80\times8.91$ & 78.4 \\
           &   & & $G$   & P200 1996 Oct &  & 0.87 & 8400 &  & \\
           &   & & ${\cal R}$ & P200 1996 Oct &  & 0.78 & 6300 &  & \\
           &   & & $I$   & P200 1997 Mar &  & 0.89 & 6600 & & \\
 CDFb      &   00:53:41.62 & $+$12:25:10.7 & $U_n$ & P200 1997/Aug 1998 Aug & 0.283 & 1.17 & 20550 & $9.05\times9.10$ & 82.4 \\
           &   & & $G$   & P200 1997 Aug/1998 Aug  &  & 1.00 & 6000 & & \\
           &   & & ${\cal R}$ & P200 1997 Aug/1998 Aug & & 0.79 & 5400 & & \\
           &   & & $I$   & P200 1998 Aug & & 1.14 & 7800 & & \\
 Q0201+1120 &  02:03:46.58 & $+$11:34:22.2   & $U_n$ & P200 1995 Nov & 0.283  & 0.95  & 27000 & $8.69\times8.72$ & 75.7 \\
            &  & & $G$   & P200 1995 Nov &        & 0.83  & 6300  &                  &       \\
            &  & & ${\cal R}$ & P200 1995 Nov &   & 0.71  & 3600  &                  &       \\
 Q0256-000 &   02:59:05.13 & $+$00:11:06.8 & $U_n$ & P200 1995 Nov/1996 Nov & 0.283 & 1.32 & 35400 & $8.54\times8.46$ & 72.2 \\
           &   & & $G$   & P200 1995 Nov/1996 Nov &       & 1.12 & 9600  &                  &      \\
           &   & & ${\cal R}$ & P200 1995 Nov/1996 Nov & & 0.91 & 6200 &                      &      \\
           &   & & $I$   & P200 1998 Sep &                & 0.95  & 9800 &                         &      \\
 Q0302-003 &   03:04:22.65 & $-$00:14:32.2 & $U_n$ & WHT 2000 Oct & 0.236 & 1.15 & 16200 & $15.59\times15.71$  & 244.9 \\ 
           &   & & $G$   & WHT 2000 Oct/KPNO 2000 Dec &  & 1.13 & 13800 &                          &       \\
           &   & & ${\cal R}$ & WHT 2000 Oct/KPNO 2000 Dec &  & 0.99 & 13800 &                     &       \\
           &   & & $I$   & KPNO 2000 Dec &   &  0.86 & 11400  &              &    \\
 B20902+34 &   09:05:31.23 & $+$34:08:01.7 & $U_n$ &  WHT 1997 Mar &  0.420 & 1.09 & 27000 & 
  $6.36\times6.57$ & 41.8 \\
           &   & & $G$ & P200/WHT 1997 Mar &       & 0.89 &  8800 &  & \\
           &   & & ${\cal R}$ & P200/WHT 1997 Mar &       & 0.88 & 4800 & &  \\
           &   & & $I$        & P200 1997 Mar &       & 0.90 & 5400 & & \\  
 Q0933+2854 &  09:33:36.09 & $+$28:45:34.8 & $U_n$ & KPNO 2000 Dec/P200 2000 Mar & 0.283 & 1.09 & 36000 & $8.93\times9.28$ & 82.9 \\
            &  & & $G$    & KPNO 2000 Dec/P200 2000 Mar &       & 0.96 & 12000 &                  &      \\
            &  & & ${\cal R}$ & KPNO 2000 Dec/P200 2000 Mar &   & 0.97 & 11000 &                  &      \\
            &  & & $I$    & KPNO 2000 Dec/P200 2000 Mar &       & 0.90 & 12600 &           &      \\
 HDF-N     &   12:36:51.31 & $+$62:13:14.3 & $U_n$  & P200 1996 Mar/Apr & 0.283 & 1.28 & 23400 & $8.62\times8.73$ & 75.3 \\
           &   & & $G$    & P200 1996 Mar/Apr &       & 1.10 & 7200  &                  &      \\
           &   & & ${\cal R}$ & P200 1996 Mar/Apr &   & 1.04 & 6000  &                  &      \\
           &   & & $I$    & P200 2000 Mar &    &  1.19 & 5400 &  & \\
 Westphal  &   14:17:43.21 & $+$52:28:48.5& $U_n$  & KPNO 1996 May & 0.471 & 1.22 & 25200 & $15.02\times15.10$ & 226.9 \\
           &   & & $G$    & KPNO 1996 May &       & 1.22 & 7200  &                    &       \\
           &   & & ${\cal R}$ & KPNO 1996 May/P200 1997 Mar & &  1.18 & 8300 &                  &       \\
 Q1422+2309 &  14:24:36.98 &  $+$22:53:49.6 & $U_n$  & WHT 1999 May & 0.236 & 0.79 & 30600 & $7.28\times15.51$ & 113.0 \\
            &  & & $G$    & WHT 1999 May &       & 0.74 & 9600  &                   &       \\
            &  & & ${\cal R}$ & WHT 1999 May &   & 0.67 & 6600 &            &       \\ 
 3C 324    &   15:49:49.75 & $+$21:28:48.1 & $U_n$  & WHT 1997 Mar/P200 1997 Mar & 0.420 & 1.12 & 23400 & $6.65\times6.63$ & 44.1 \\
           &   & & $G$    & WHT 1997 Mar &       & 1.09 & 7200 &                  &      \\
           &   & & ${\cal R}$ & WHT 1997 Mar &   & 1.14 & 4800 &                  &      \\
           &   & & $I$    & P200 1997 Mar &      & 0.88 & 3600 &                  &      \\  
 SSA22a    &   22:17:33.82 & $+$00:15:04.4 & $U_n$  & P200 1995 Aug/1996 Aug & 0.283 & 1.03 & 26000 & $8.74\times8.89$ & 77.7 \\
           &   & & $G$    & P200 1995 Aug/1996 Aug &       & 1.02 & 6300  &                  &      \\
           &   & & ${\cal R}$ & P200 1995 Aug/1996 Aug &   & 1.04 & 6000  &                  &      \\
           &   & & $I$    & P200 1998 Aug &                & 0.89 & 7200  &                  &      \\
 SSA22b    &   22:17:33.83 & $+$00:06:21.8 & $U_n$  & P200 1996 Aug/Oct & 0.283 & 1.08 & 33050 & $8.64\times8.98$ & 77.6 \\
           &   & & $G$    & P200 1996 Aug/Oct &       & 0.90 & 8100  &                  &      \\
           &   & & ${\cal R}$ & P200 1996 Aug/Oct &   & 0.90 & 8100  &                  &      \\
           &   & & $I$    & P200 1998 Aug  &          & 0.86 & 6600  &                  &      \\
 DSF2237a  &   22:40:08.32 & $+$11:52:41.1 & $U_n$  & P200 1997 Aug/1998 Aug & 0.283 & 0.96 & 23400 & $9.08\times9.08$ & 83.4 \\
           &   & & $G$    & P200 1997 Aug/1998 Aug &       & 0.99 & 6900  &                  &      \\
           &   & & ${\cal R}$ & P200 1997 Aug/1998 Aug &   & 0.94 & 6300  &                  &      \\
           &   & & $I$    & P200 1998 Aug &                & 1.04 & 7200  &                  &      \\
 DSF2237b  &   22:39:34.10 & $+$11:51:38.8 & $U_n$  & P200 1997 Aug & 0.283  & 0.98 & 23400 & $8.99\times9.08$  & 81.7 \\
           &   & & $G$    & P200 1997 Aug &        & 0.86 & 8400  &                   &      \\
           &   & & ${\cal R}$ & P200 1997 Aug &    & 0.84 & 6900  &                   &      \\
           &   & & $I$    & P200 1998 Aug &        & 0.81 & 6000  &                   &      \\
 Q2233+1341 &  22:36:08.76 & $+$13:56:22.2 & $U_n$  & P200 1999 Aug/Oct & 0.283 & 1.18 & 20820 & $9.25\times9.25$ & 85.6 \\
            &  & & $G$    & P200 1999 Aug/Oct &       & 1.18 & 15600 &                  &      \\
            &  & & ${\cal R}$ & P200 1999 Aug/Oct &   & 1.09 & 14100 &                  &      \\
            &  & & $I$    & P200 1999 Oct    &        & 0.95 & 16200 &                  &      \\
\enddata
\tablenotetext{a}{Positions of the field centers.}
\tablenotetext{b}{P200: Palomar 5.08m telescope ; WHT: William Herschel 4.2m telescope ; LCO: Las Campanas Observatory
DuPont 2.5m telescope ; NTT: ESO 3.6m New Technology Telescope; KPNO: Kitt Peak 4m Mayall telescope.} 
\end{deluxetable}

\begin{deluxetable}{lccccccc}
\tabletypesize{\scriptsize}
\tablewidth{0pc}
\tablecaption{Photometry}
\tablehead{
\colhead{Field Name} & \colhead{E(B$-$V)} & \colhead{$\sigma(U_n)$\tablenotemark{a}} & \colhead{$\sigma(G)$\tablenotemark{a}}  & \colhead{$\sigma({\cal R})$\tablenotemark{a}} & \colhead{$\langle N \rangle$\tablenotemark{b}} & 
\colhead{$\langle G-{\cal R} \rangle$\tablenotemark{c}} & \colhead{$\langle U_n-G \rangle$\tablenotemark{c}} 
} 
\startdata
Q0000$-$263 & 0.000 & 28.80 & 28.64 & 28.06 & $29.3\pm1.4$ & $0.81\pm0.02$ & $0.46\pm0.02$ \\
CDFa & 0.040 & 28.37 & 28.72 & 28.16 & $25.8\pm0.6$ & $0.76\pm0.01$ & $0.51\pm0.01$ \\
CDFb & 0.040 & 28.23 & 28.47 & 27.93 & $22.9\pm0.6$ & $0.77\pm0.01$ & $0.47\pm0.01$ \\
Q0201$+$113 & 0.142 & 28.16 & 28.47 & 27.86 & $29.2\pm0.6$ & $0.79\pm0.01$ & $0.58\pm0.01$ \\
Q0256$-$000 & 0.085 & 28.41 & 28.34 & 27.96 & $26.7\pm0.6$ & $0.77\pm0.01$ & $0.63\pm0.01$ \\
Q0302$-$003 & 0.089 & 28.02 & 28.66 & 28.14 & $25.2\pm0.3$ & $0.68\pm0.01$ & $0.68\pm0.01$ \\
B20902$+$34 & 0.025 & 28.43 & 28.58 & 27.73 & $26.9\pm0.8$ & $0.70\pm0.02$ & $0.54\pm0.02$ \\
Q0933$+$289 & 0.023 & 28.35 & 28.82 & 28.31 & $23.5\pm0.6$ & $0.74\pm0.01$ & $0.57\pm0.01$ \\
HDF-N & 0.000 & 27.84 & 28.41 & 27.98 & $25.5\pm0.6$ & $0.76\pm0.01$ & $0.53\pm0.01$ \\
Westphal & 0.000 & 28.32 & 28.11 & 27.33 & $25.0\pm0.3$ & $0.82\pm0.01$ & $0.55\pm0.01$ \\
Q1422$+$2309 & 0.032 & 29.42 & 29.36 & 28.49 & $29.4\pm0.4$ & $0.80\pm0.01$ & $0.65\pm0.01$ \\
3C324 & 0.039 & 28.24 & 28.49 & 27.67 & $23.7\pm0.8$ & $0.76\pm0.02$ & $0.59\pm0.02$ \\
SSA22a & 0.080 & 28.11 & 28.13 & 27.78 & $25.5\pm0.6$ & $0.74\pm0.01$ & $0.56\pm0.01$ \\
SSA22b & 0.080 & 28.16 & 28.53 & 28.26 & $25.1\pm0.6$ & $0.75\pm0.01$ & $0.47\pm0.01$ \\
DSF2237a & 0.048 & 28.36 & 28.54 & 28.08 & $26.2\pm0.6$ & $0.78\pm0.01$ & $0.56\pm0.01$ \\
DSF2237b & 0.048 & 28.53 & 29.00 & 28.47 & $24.6\pm0.6$ & $0.70\pm0.01$ & $0.62\pm0.01$ \\
Q2233$+$1341 & 0.048 & 28.19 & 28.50 & 27.96 & $25.5\pm0.6$ & $0.85\pm0.01$ & $0.57\pm0.01$ \\
 & & & & & & & \\
Average\tablenotemark{d} & & & & & $25.9\pm1.9$ & $0.76\pm0.04$ & $0.56\pm0.06$ \\
\enddata
\tablenotetext{a}{One sigma limits for objects having an isophotal size equivalent
to 3 times the seeing disk (see text).}
\tablenotetext{b}{Mean number of detected galaxies per square arc minute in the ${\cal R}$
magnitude range $22.5-25.0$. This statistic provides an estimate of field-to-field
fluctuations that should be largely independent of depth and seeing among the fields
surveyed.}
\tablenotetext{c}{Mean color of galaxies significantly detected in $U_n$ and$G$ to
the survey limit in ${\cal R}$. This statistic provides an idea of the consistency in
the galaxy colors from field to field (see text). }
\tablenotetext{d}{Unweighted average values from field to field, where the quoted
uncertainties are the field-to-field scatter in the mean quantities.}

\end{deluxetable}
\newpage

\begin{deluxetable}{lccccc}
\tabletypesize{\scriptsize}
\tablewidth{0pc}
\tablecaption{Typical Photometric Uncertainties and Biases}
\tablehead{
\colhead{${\cal R}$ Range\tablenotemark{a}} & \colhead{$G-{\cal R}$ Range\tablenotemark{b}} & \colhead{$\Delta{\cal R}$\tablenotemark{c}} & 
\colhead{$\sigma({\cal R})$\tablenotemark{d}} &
\colhead{$\Delta(G-{\cal R})$\tablenotemark{e}} & \colhead{$\sigma(G-{\cal R})$\tablenotemark{f}} }
\startdata
22.50--23.00 & 0.00--0.20 & $-$0.03 & 0.09 & ~0.02 & ~0.04  \\
   & 0.20--0.40 & $-$0.06 & 0.10 & ~0.02 & ~0.05   \\
   & 0.40--0.60 & $-$0.08 & 0.12 & ~0.03 & ~0.06   \\
   & 0.60--0.80 & $-$0.08 & 0.12 & ~0.03 & ~0.08   \\
   & 0.80--1.00 & $-$0.08 & 0.12 & ~0.05 & ~0.08   \\
   & 1.00--1.20 & $-$0.08 & 0.12 & ~0.04 & ~0.11   \\
23.00--23.50 & 0.00--0.20 & $-$0.07 & 0.12 & 0.02 & ~0.04   \\
   & 0.20--0.40 & $-$0.07 & 0.12 & ~0.02 & ~0.05   \\
   & 0.40--0.60 & $-$0.07 & 0.12 & ~0.02 & ~0.07   \\
   & 0.60--0.80 & $-$0.07 & 0.12 & ~0.03 & ~0.07   \\
   & 0.80--1.00 & $-$0.08 & 0.12 & ~0.05 & ~0.08   \\
   & 1.00--1.20 & $-$0.08 & 0.12 & ~0.05 & ~0.12   \\
23.50--24.00 & 0.00--0.20 & $-$0.08 & 0.14 & ~0.02 & ~0.07   \\
   & 0.20--0.40 & $-$0.08 &  0.14 & ~0.01 & ~0.07   \\
   & 0.40--0.60 & $-$0.08 & 0.14 & ~0.03 & ~0.08    \\
   & 0.60--0.80 & $-$0.08 & 0.14 & ~0.04 & ~0.09   \\
   & 0.80--1.00 & $-$0.08 & 0.14 & ~0.05 & ~0.08   \\
   & 1.00--1.20 & $-$0.09 & 0.15 & ~0.05 & ~0.12   \\
24.00--24.50 & 0.00--0.20 & $-$0.07 & 0.18 & ~0.00 & ~0.10   \\
   & 0.20--0.40 & $-$0.07 &  0.18 & ~0.01 & ~0.10   \\
   & 0.40--0.60 & $-$0.07 & 0.18 & ~0.03 & ~0.11   \\
   & 0.60--0.80 & $-$0.07 & 0.18 & ~0.05 & ~0.13   \\
   & 0.80--1.00 & $-$0.07 & 0.18 & ~0.08 & ~0.15   \\
   & 1.00--1.20 & $-$0.08 & 0.19 & ~0.09 & ~0.16   \\
24.50--25.00 & 0.00--0.20 & $-$0.02 &  0.21 & $-$0.03 & ~0.13   \\
   & 0.20--0.40 & $-$0.02 & 0.21 & ~0.00 & ~0.13   \\
   & 0.40--0.60 & $-$0.02 & 0.21 & ~0.03 & ~0.13   \\
   & 0.60--0.80 & $-$0.02 & 0.21 & ~0.06 & ~0.16   \\
   & 0.80--1.00 & $-$0.02 & 0.21 & ~0.10 & ~0.17   \\
   & 1.00--1.20 & $-$0.02 & 0.21 & ~0.13 & ~0.20   \\
25.00--25.50 & 0.00--0.20 & $-$0.02 & 0.23 & $-$0.07 & ~0.15   \\
   & 0.20--0.40 & $-$0.02 & 0.23 & $-$0.03 & ~0.16   \\
   & 0.40--0.60 & $-$0.02 & 0.23 & ~0.01 & ~0.17   \\
   & 0.60--0.80 & $-$0.02 & 0.23 & ~0.07 & ~0.19   \\
   & 0.80--1.00 & $-$0.02 & 0.23 & ~0.13 & ~0.20   \\
   & 1.00--1.20 & $-$0.02 & 0.25 & ~0.16 & ~0.24   \\
\enddata
\tablenotetext{a}{Recovered magnitude range for artificial galaxies satisfying the
Lyman break galaxy selection criteria.}
\tablenotetext{b}{Recovered color range for artificial galaxies satisfying the LBG
selection 
criteria.}
\tablenotetext{c}{The average value of ${\cal R}_{\rm meas}-{\cal R}_{\rm true}$. Significant departures
from zero imply significant biases in the photometry.} 
\tablenotetext{d}{The RMS of the difference between recovered ${\cal R}$ magnitudes and input magnitudes.}
\tablenotetext{e}{The average value of $(G-{\cal R})_{\rm meas} - (G-{\cal R})_{\rm true})$ for objects
recovered as LBG candidates}. 
\tablenotetext{f}{The RMS of the difference between recovered $G-{\cal R}$ colors and the true input colors.}
\end{deluxetable}

\newpage

\begin{deluxetable}{ll}
\tabletypesize{\scriptsize}
\tablewidth{0pc}
\tablecaption{Lyman Break Galaxy Photometric Selection Criteria}
\tablehead{
\colhead{Candidate Type} & \colhead{Criteria} 
} 
\startdata
 C &       ${\cal R}\le25.5 ; \quad (G-{\cal R})\le 1.2; \quad (U_n-G) > (G-{\cal R})+1.5$; undetected in $U_n$ \\

 D &       ${\cal R}\le25.5 ; \quad (G-{\cal R})\le 1.2;\quad (U_n-G) > (G-{\cal R})+1.5$; detected in $U_n$ \\

 M &      ${\cal R}\le25.5; \quad (G-{\cal R})\le 1.2; \quad 1.0\le (U_n-G)-(G-{\cal R}) \le 1.5$; undetected in $U_n$ \\

 MD &       ${\cal R}\le25.5; \quad (G-{\cal R})\le 1.2; \quad 1.0\le (U_n-G)-(G-{\cal R}) \le 1.5$; detected in $U_n$ \\
\enddata
\end{deluxetable}

\begin{deluxetable}{lccccccc}
\tabletypesize{\scriptsize}
\tablewidth{0pc}
\tablecaption{Spectroscopic Observations}
\tablehead{
\colhead{Field Name} & \colhead{N(Cand)\tablenotemark{a}} & \colhead{N(Obs)\tablenotemark{b}} & 
\colhead{N($z>2$)\tablenotemark{c}}  & \colhead{N(Int)\tablenotemark{d}} &
\colhead{$f_{obs}$\tablenotemark{e}} & \colhead{$f_{LBG}$\tablenotemark{f}} & 
\colhead{$f_{int}$\tablenotemark{g}}
} 
\startdata
Q0000$-$263 & ~28 & ~22 & ~19 & 1 & 0.79 & 0.86 & 0.05 \\
CDFa & 100 & ~60 & ~40 & 1 &  0.60 & 0.67 & 0.02 \\
CDFb & 121 & ~44 & ~29 & 0 &  0.36 & 0.66 & 0.00 \\
Q0201$+$113 & ~87 & ~32 & 21 & 0 &  0.37 & 0.66 & 0.00 \\
Q0256$-$000 & 120 & ~76 & 55 & 1 & 0.63 & 0.72 & 0.06 \\
Q0302$-$003 & 191 & ~94 & 50 & 3 & 0.49 & 0.53 & 0.02 \\
B20902$+$34 & ~78 & ~46 & 38 & 2 &  0.59 & 0.83 & 0.05 \\
Q0933$+$289 & 211 & 129 & 76 & 4 &  0.61 & 0.59 & 0.02 \\
HDF-N & 132 & ~68 & ~47 & 0 &  0.52 & 0.69 & 0.00 \\
Westphal & 329 & 232 & 188 & 7 & 0.71 & 0.81 & 0.04 \\
Q1422$+$2309 & 273 & 167 & 120 & 5 & 0.61 & 0.72 & 0.04 \\
~~(${\cal R}\le26.0)$)\tablenotemark{h}  & 453 & 195 & 135 & 5 & 0.43 & 0.69 &  0.04 \\
3C324 & ~51 & ~12 & ~11 & 0 & 0.24 & 0.92 & 0.00 \\
SSA22a & 146 & ~72 & ~57 & 6 &  0.49 & 0.79 & 0.10 \\
SSA22b & ~89 & ~48 & ~34 & 1 & 0.54 & 0.71 & 0.03 \\
DSF2237a & 121 & ~57 & ~44 & 5 & 0.47 & 0.77 & 0.10 \\
DSF2237b & 176 & ~68 & ~58 & 4 & 0.39 & 0.85 & 0.06 \\
Q2233$+$1341 & ~94 & ~66 & ~53 & 5 & 0.70 & 0.80 & 0.09 \\
             &    &    &    &   &      &      &      \\
TOTAL\tablenotemark{i} & 2347 & 1293 & 940 & 45 & 0.55 & 0.73 & 0.05 \\
\enddata
\tablenotetext{a}{Number of LBG candidates in field.}
\tablenotetext{b}{Number of LBG candidates observed spectroscopically.}
\tablenotetext{c}{Number of spectroscopically identified LBG candidates with $z>2$.}.
\tablenotetext{d}{Number of spectroscopically identified ``interlopers''. Of the total
of 45, 40 are stars, 2 are galaxies with $z\simeq1.98$, and 3 are absorption line galaxies
at $z \sim 0.5$ (see text).}
\tablenotetext{e}{The fraction of LBG candidates observed spectroscopically.}
\tablenotetext{f}{The fraction of spectroscopically observed candidates with measured $z > 2$.}
\tablenotetext{g}{The fraction of ``interlopers'' among spectroscopically identified LBG candidates.}
\tablenotetext{h}{For objects with ${\cal R} \le 26.0$; all other catalogs are limited at ${\cal R}\le 25.5$.}
\tablenotetext{i}{The total numbers include only candidates with ${\cal R}\le 25.5$, and
so the numbers do not include the 28 objects (15 of which yielded successful redshifts) fainter
than ${\cal R}=25.5$ observed in the Q1422$+$2309 field.} 
\end{deluxetable}
\newpage

\begin{deluxetable}{lccccccc}
\tabletypesize{\scriptsize}
\tablewidth{0pc}
\tablecaption{Redshift Statistics for LBG Candidates with ${\cal R}\le 25.5$}
\tablehead{
\colhead{Candidate Type} & \colhead{N(Cand)\tablenotemark{a}} & \colhead{N(obs)\tablenotemark{b}}
& \colhead{N($z>2$)\tablenotemark{c}}& \colhead{$f_{obs}$\tablenotemark{d}} 
& \colhead{$f_{LBG}$\tablenotemark{e}} & \colhead{$f_{Int}$\tablenotemark{f}} &  \colhead{$\langle z \rangle$\tablenotemark{g}}}  
\startdata
C & 526 & 366 & 291 & 0.70 & 0.80 & 0.03 & $3.09\pm 0.22$ \\
D & 343 & 237 & 197 & 0.69 & 0.83 & 0.02 & $2.93\pm 0.26$ \\
M & 449 & 197 & 136 & 0.44 & 0.69 & 0.02 & $3.15\pm 0.24$ \\
MD & 1029 & 486 & 316 & 0.47 & 0.65 & 0.08 & $2.79\pm 0.27$ \\\\
Total & 2347 & 1286 & 940 & 0.55 & 0.73 & 0.05 & $2.96\pm 0.29$ \\\\
S99\tablenotemark{h} & 1892 & 1104 & 822 & 0.58 & 0.74 & 0.05 & $3.01\pm 0.27$ \\ 
\enddata
\tablenotetext{a}{Number of photometric candidates.}    
\tablenotetext{b}{Number of objects spectroscopically observed.}
\tablenotetext{c}{Number of objects with confirmed redshifts $z>2$.}
\tablenotetext{d}{The fraction of candidates observed spectroscopically.}
\tablenotetext{e}{The fraction of spectroscopically observed candidates yielding confirmed
redshifts $z>2$.}
\tablenotetext{f}{The fraction of spectroscopically identified objects that have $z < 2$.}
\tablenotetext{g}{Mean and standard deviation of the redshift distribution for objects
with $z>2$.}
\tablenotetext{h}{Number of objects satisfying the photometric criteria used in Steidel \et 1999.
These criteria exclude objects with $U_n-G < 1.6$.}
\end{deluxetable}

\newpage

\begin{deluxetable}{lcccccrrcc}
\tabletypesize{\scriptsize}
\tablewidth{0pc}
\tablecaption{Q0000$-$263 Field LBGs}
\tablehead{
\colhead{Name} & \colhead{$\alpha(J2000)$} & \colhead{$\delta(J2000)$} & \colhead{${\cal R}$} &
\colhead{$G-{\cal R}$} & \colhead{$U_n-G$} & \colhead{${\rm z_{em}}$} & \colhead{${\rm z_{abs}}$} &
\colhead{Type} & \colhead{Notes\tablenotemark{a}}} 
\startdata
Q0000-C1 & 00:03:17.13 & -26:06:03.2 & 25.38 & 0.60 & 2.86 & -1.000 & -1.000 & --- &    \\
Q0000-C2 & 00:03:30.38 & -26:05:47.3 & 25.23 & 0.65 & 3.07 & -1.000 & -1.000 & --- &    \\
Q0000-C3 & 00:03:27.58 & -26:05:45.0 & 23.99 & 1.17 & 3.50 & 0.000 & 0.000 & STAR &  S96-C02  \\
Q0000-C4 & 00:03:21.87 & -26:05:39.9 & 24.57 & 1.06 & 2.87 & 3.594 & 3.580 & GAL &    \\
Q0000-C5 & 00:03:28.94 & -26:05:26.1 & 23.62 & 1.08 & 4.12 & 3.791 & -2.000 & QSO &  S96-C04  \\
Q0000-C6 & 00:03:21.12 & -26:04:17.4 & 25.22 & 1.00 & 2.73 & -2.000 & 3.389 & GAL &  S96-C07  \\
Q0000-C7 & 00:03:28.85 & -26:03:53.3 & 24.21 & 0.07 & 4.40 & 3.426 & -2.000 & AGN &  S96-C09; G2\tablenotemark{b}  \\
Q0000-C8 & 00:03:20.25 & -26:03:36.6 & 24.47 & 0.89 & 3.32 & -2.000 & :3.427 & GAL &  S96-C10  \\
Q0000-C9 & 00:03:26.33 & -26:03:28.7 & 25.03 & 1.02 & 2.78 & -1.000 & -1.000 & --- &    \\
Q0000-C10 & 00:03:23.30 & -26:02:52.3 & 24.78 & 0.84 & 3.08 & -2.000 & 2.960 & GAL &  S96-C13  \\
Q0000-C12 & 00:03:27.31 & -26:02:27.7 & 24.37 & 0.63 & 3.56 & 3.021 & -2.000 & GAL &  S96-C14  \\
Q0000-C13 & 00:03:25.40 & -26:01:25.3 & 25.03 & 0.76 & 2.84 & -2.000 & -2.000 & --- &    \\
Q0000-C14 & 00:03:30.39 & -26:01:20.7 & 24.47 & 0.86 & 3.24 & 3.057 & 3.053 & AGN &  S96-C16\\
Q0000-D1 & 00:03:25.17 & -26:04:35.2 & 24.32 & 0.49 & 2.20 & 2.788 & -2.000 & GAL &  S96-C22  \\
Q0000-D2 & 00:03:31.02 & -26:04:31.1 & 25.23 & 0.34 & 2.45 & -1.000 & -1.000 & --- &    \\
Q0000-D3 & 00:03:17.23 & -26:04:03.9 & 24.91 & 0.51 & 3.33 & 3.143 & 3.147 & GAL &  S96-C08  \\
Q0000-D4 & 00:03:20.97 & -26:03:37.6 & 24.79 & 0.47 & 2.09 & -2.000 & 2.775 & GAL &  S96-C27  \\
Q0000-D5 & 00:03:28.51 & -26:03:28.3 & 24.96 & 0.54 & 2.23 & 3.150 & 3.139 & GAL &  S96-C11  \\
Q0000-D6 & 00:03:23.79 & -26:02:48.5 & 22.88 & 0.45 & 2.26 & 2.971 & 2.958 & GAL &    \\
Q0000-D8 & 00:03:32.45 & -26:01:47.4 & 24.76 & 0.62 & 2.64 & -2.000 & 3.041 & GAL &  S96-C25  \\
Q0000-M1 & 00:03:19.84 & -26:01:22.1 & 25.05 & 1.03 & 2.42 & -2.000 & 3.168 & GAL &  S96-C17  \\
Q0000-MD1 & 00:03:24.65 & -26:05:48.8 & 25.38 & 0.72 & 2.25 & 3.602 & 3.589 & GAL &    \\
Q0000-MD2 & 00:03:25.12 & -26:04:52.6 & 24.88 & 1.11 & 2.28 & 3.596 & -2.000 & GAL &  S96-C05  \\
Q0000-MD3 & 00:03:18.47 & -26:04:35.0 & 23.75 & 0.54 & 1.76 & -2.000 & 3.201 & GAL &  S96-C23  \\
Q0000-MD4 & 00:03:19.70 & -26:04:26.5 & 24.82 & 0.49 & 1.70 & -1.000 & -1.000 & --- &    \\
Q0000-MD5 & 00:03:25.92 & -26:02:20.0 & 25.28 & 0.40 & 1.92 & -2.000 & -2.000 & --- &    \\
Q0000-MD6 & 00:03:21.22 & -26:01:47.6 & 25.41 & 0.66 & 2.03 & 3.092 & -2.000 & GAL &    \\
Q0000-MD7 & 00:03:21.25 & -26:01:26.8 & 22.68 & 0.74 & 1.76 & -1.000 & -1.000 & --- &    \\
\enddata
\tablenotetext{a}{"S96-XXX" indicates object names appearing in Steidel \et 1996a}    
\tablenotetext{b}{This object is discussed in Macchetto \et 1993 and Giavalisco \et 1994,1995}
\end{deluxetable}

\input tab8_short.tex
\input tab9_short.tex
\input tab10_short.tex
\clearpage
\input tab11_short.tex
\input tab12_short.tex
\input tab13_short.tex
\input tab14_short.tex 
\input tab15_short.tex
\input tab16_short.tex
\input tab17_short.tex
\clearpage
\input tab18_short.tex
\input tab19_short.tex
\input tab20_short.tex
\input tab21_short.tex
\input tab22_short.tex
\input tab23_short.tex
\newpage
\begin{figure}
\plotone{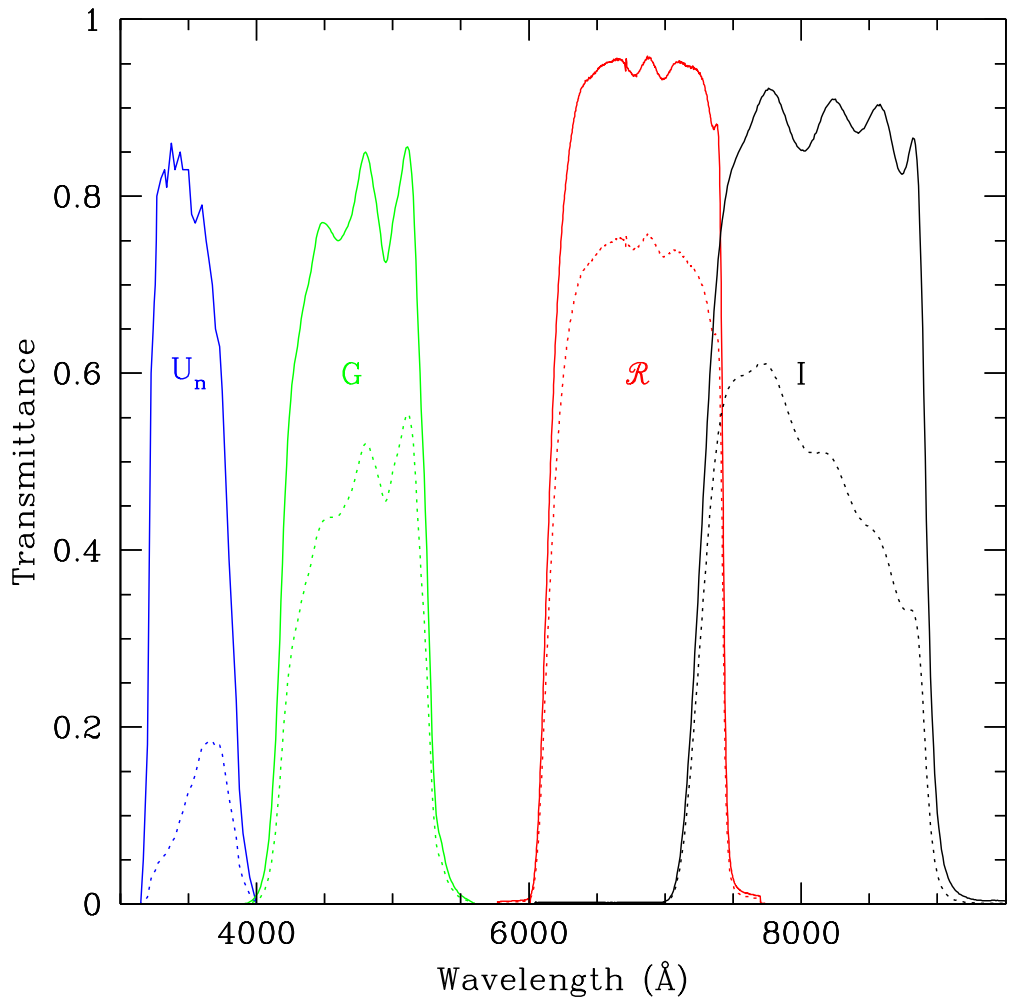}
\caption{The filter system used for the deep imaging, $U_n$ [3550/600], $G$ [4780/1100],
${\cal R}$ [6830/1250], and $I$ [8100/1650]. The dotted curves show the effective bandpasses
including typical CCD quantum efficiency and atmospheric attenuation. }
\end{figure}
\newpage
\begin{figure}
\plotone{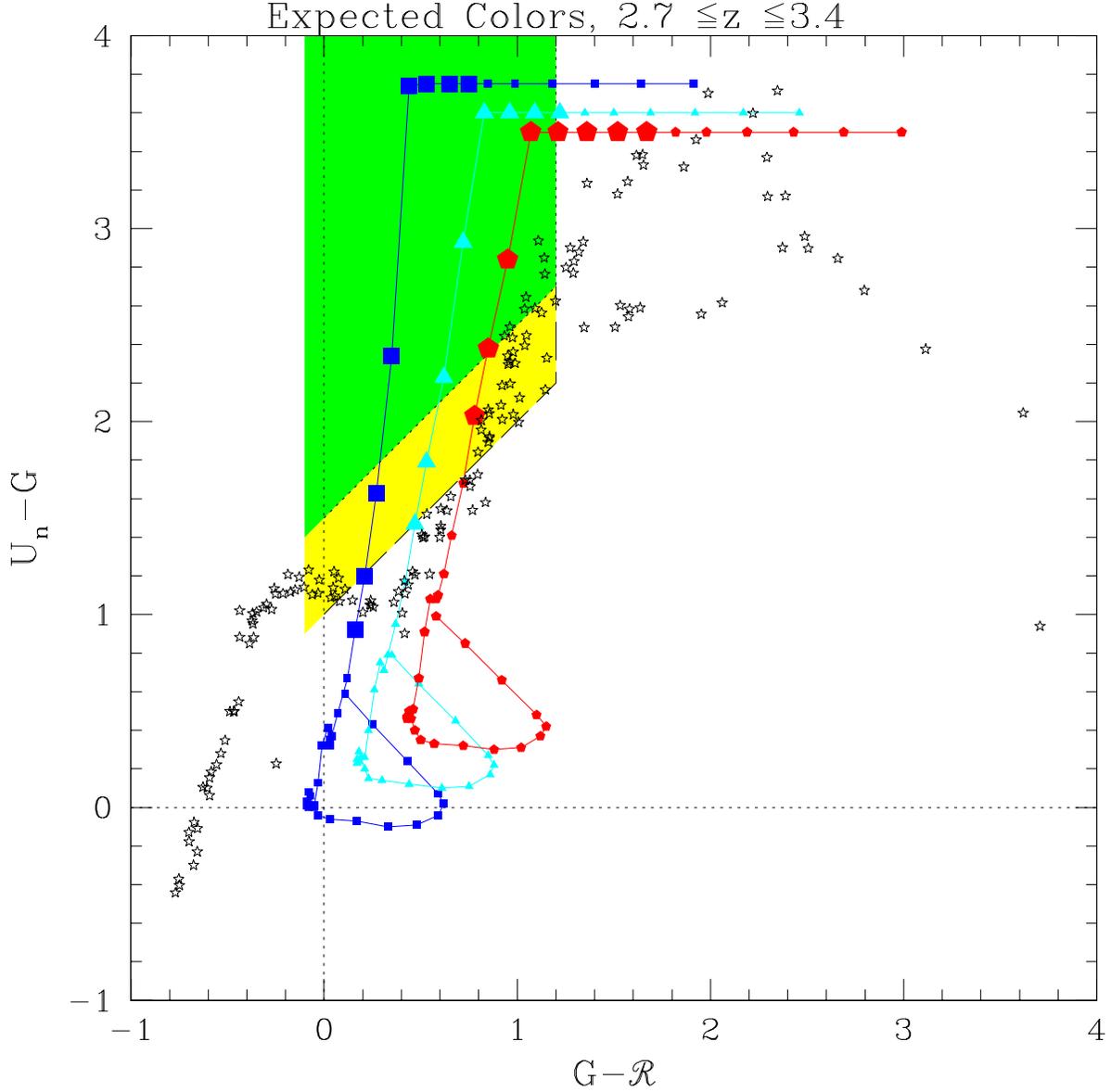}
\caption{Plot showing the expected colors for model star forming 
galaxies in the targeted redshift range,
for 3 assumed values of internal extinction (${\rm E(B-V)=0,0.15,0.30}$ using the
Calzetti \et 2000 prescription, for squares,
triangles, and pentagons, respectively), with points corresponding to intervals
of $\Delta z =0.1$. The large points on each curve correspond to galaxies in the
redshift interval $2.7 \le z \le 3.4$. The $U_n-G$ colors have been truncated for clarity;
in practice, limited dynamic range will prevent $U_n-G$ limits from exceeding
$\sim 4.0$. The green shaded region is where ``C'' and ``D''
type candidates are located; the yellow shaded region corresponds to candidate types ``M'' and ``MD''
(see Table 4). Note that the redshift distributions of LBG samples are expected to depend upon
the intrinsic spectral shape of the galaxies, which we have parameterized by E(B-V); conversely,
the range of intrinsic colors sampled by the selection 
criteria will depend upon redshift. These points are 
discussed in some detail in Steidel \et 1999. 
The expected location of the stellar locus, based on the Gunn \& Stryker (1983) atlas, is shown with the black ``stars''.}
\end{figure}
\begin{figure}
\plotone{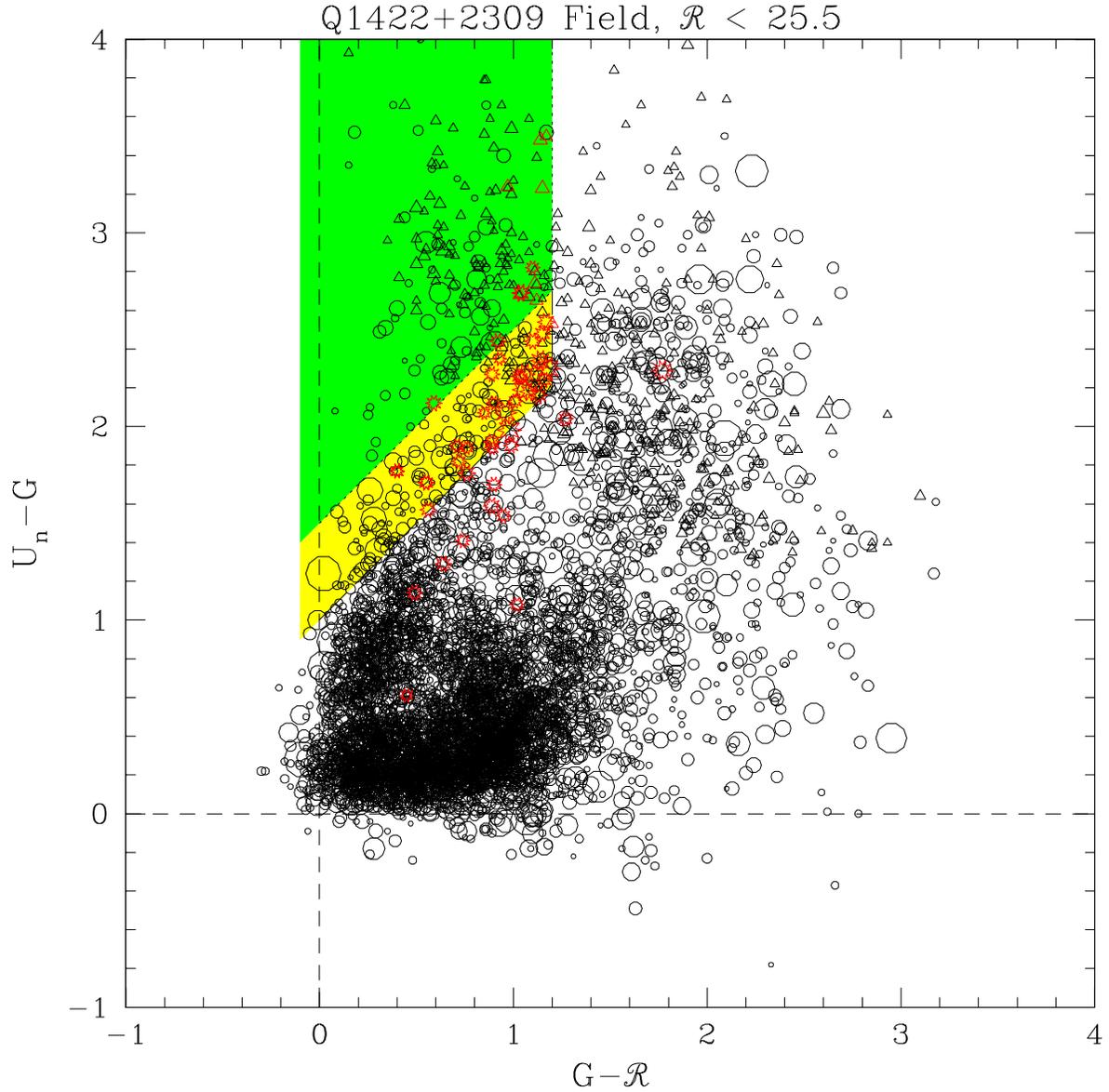}
\caption{Plot showing real data, for ${\cal R}\le25.5$ objects in the Q1422+2309 field. Objects that
are detected in $U_n$ are shown with open circles (scaled according to brightness); objects having
only limits on $U_n-G$ colors are shown with triangles. The locations
in the two-color plane of all 55 stars (taken from all 17 of the observed fields) that have been observed spectroscopically (40 of which satisfy
the LBG color criteria) are shown in red.} 
\end{figure}
\newpage
\begin{figure}
\plotone{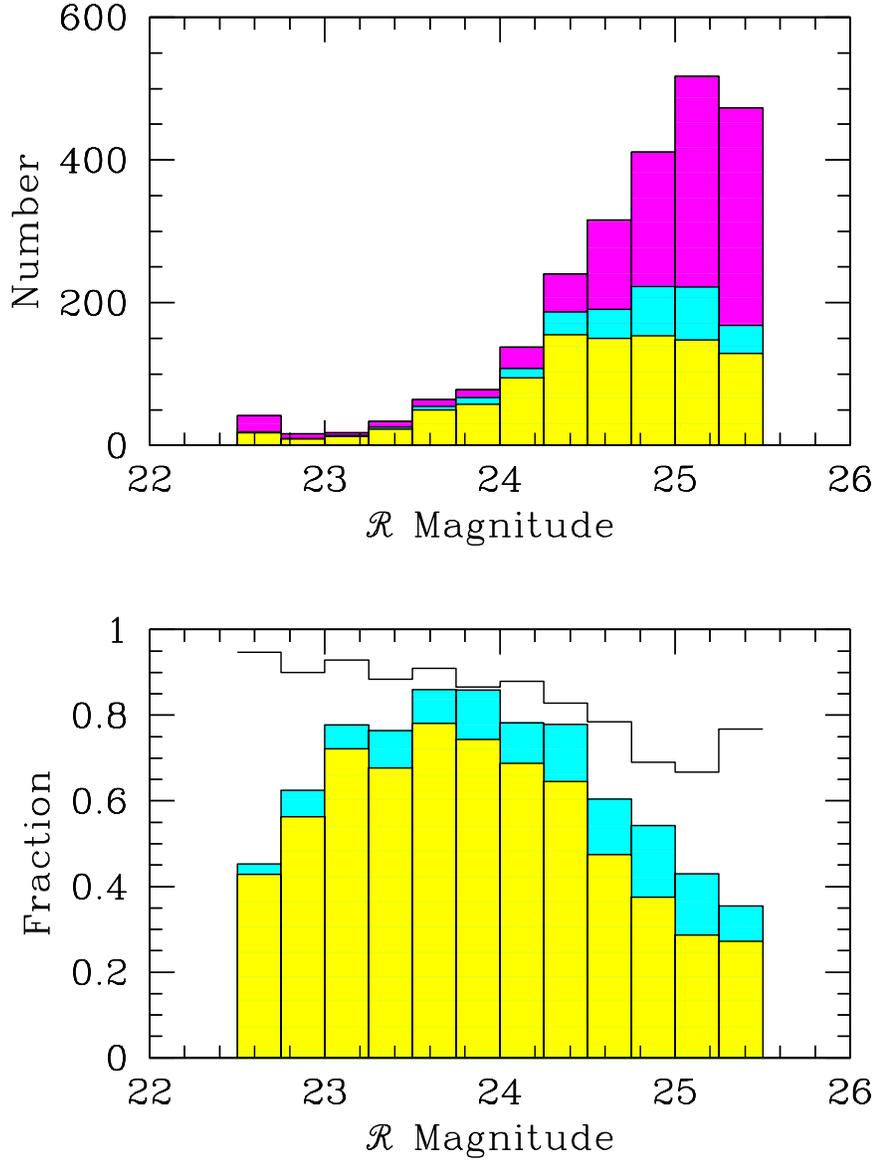}
\caption{Plot showing the magnitude distribution of the full photometric LBG sample (magenta),
the sub-sample targeted for spectroscopy (cyan), and the subsample with successful spectroscopic
redshifts (yellow). The bottom panel shows the fraction of  
the full photometric sample represented by the spectroscopically observed subsample (cyan) and the spectroscopically
successful samples versus apparent magnitude. The black curve represents the fraction of spectroscopically observed
objects that yielded successful redshifts, as a function of apparent ${\cal R}$ magnitude. } 
\end{figure}
\begin{figure}
\plotone{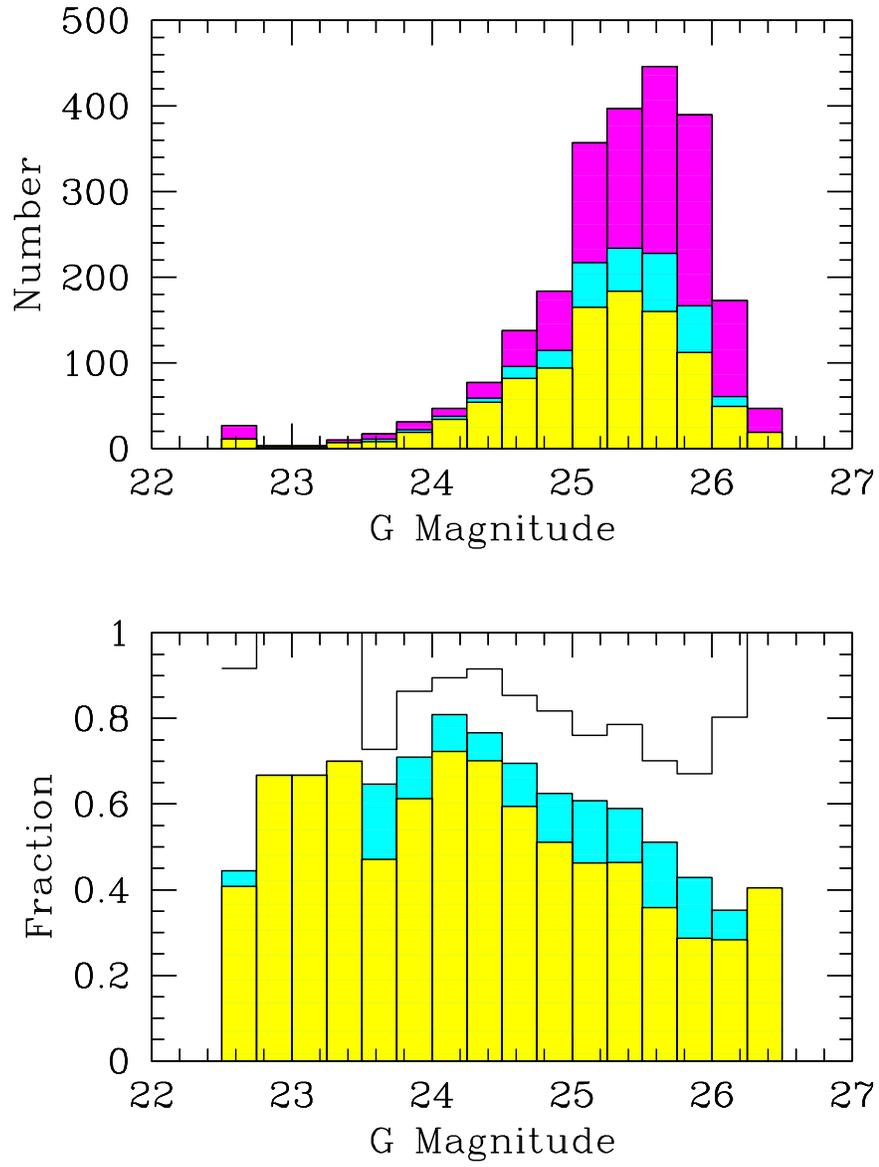}
\caption{Same as Figure 4, as a function of $G$ magnitude.}
\end{figure}
\begin{figure}
\plotone{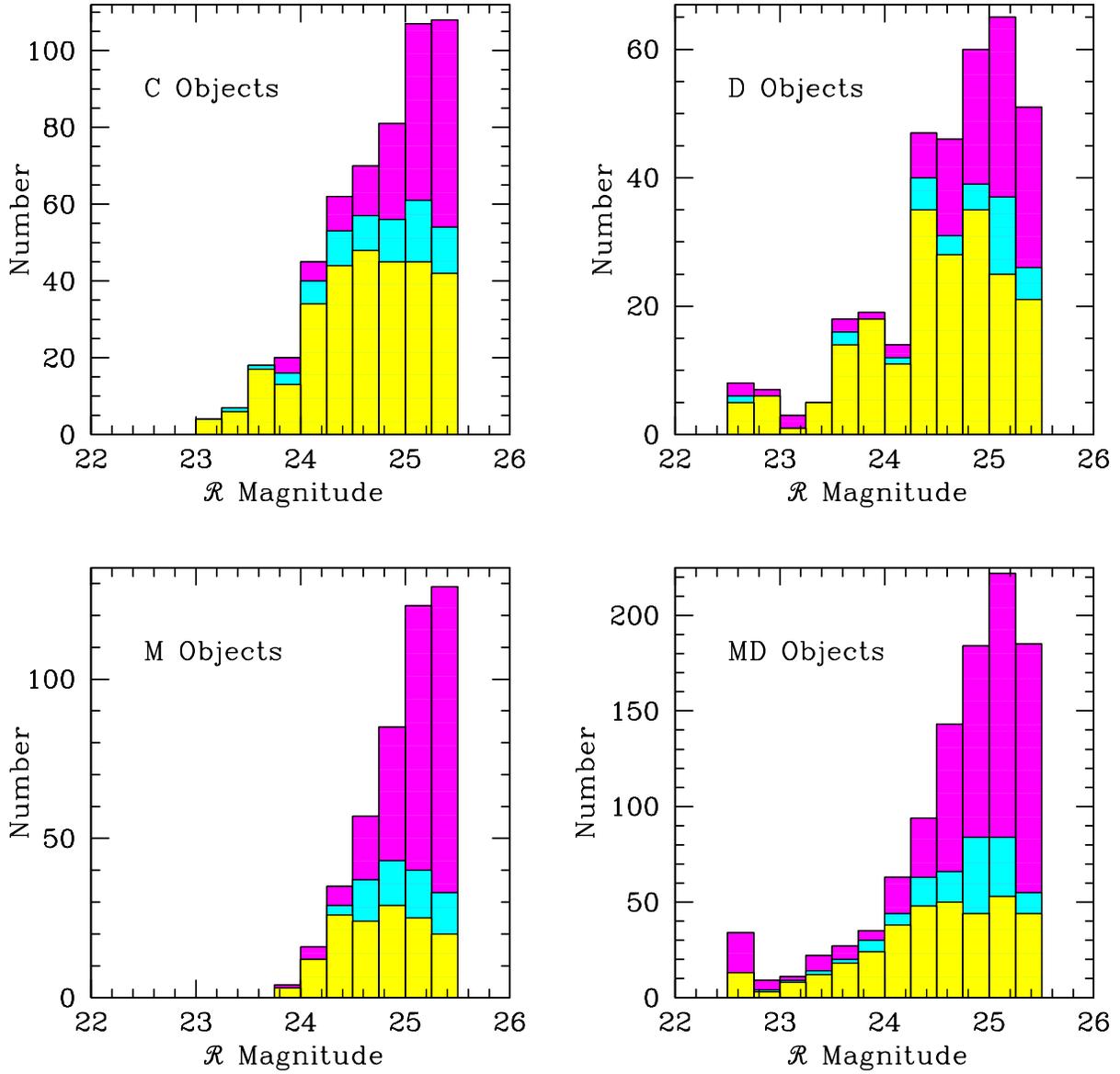}
\caption{Same as figure 4, divided by candidate classification system.}
\end{figure}
\begin{figure}
\plotone{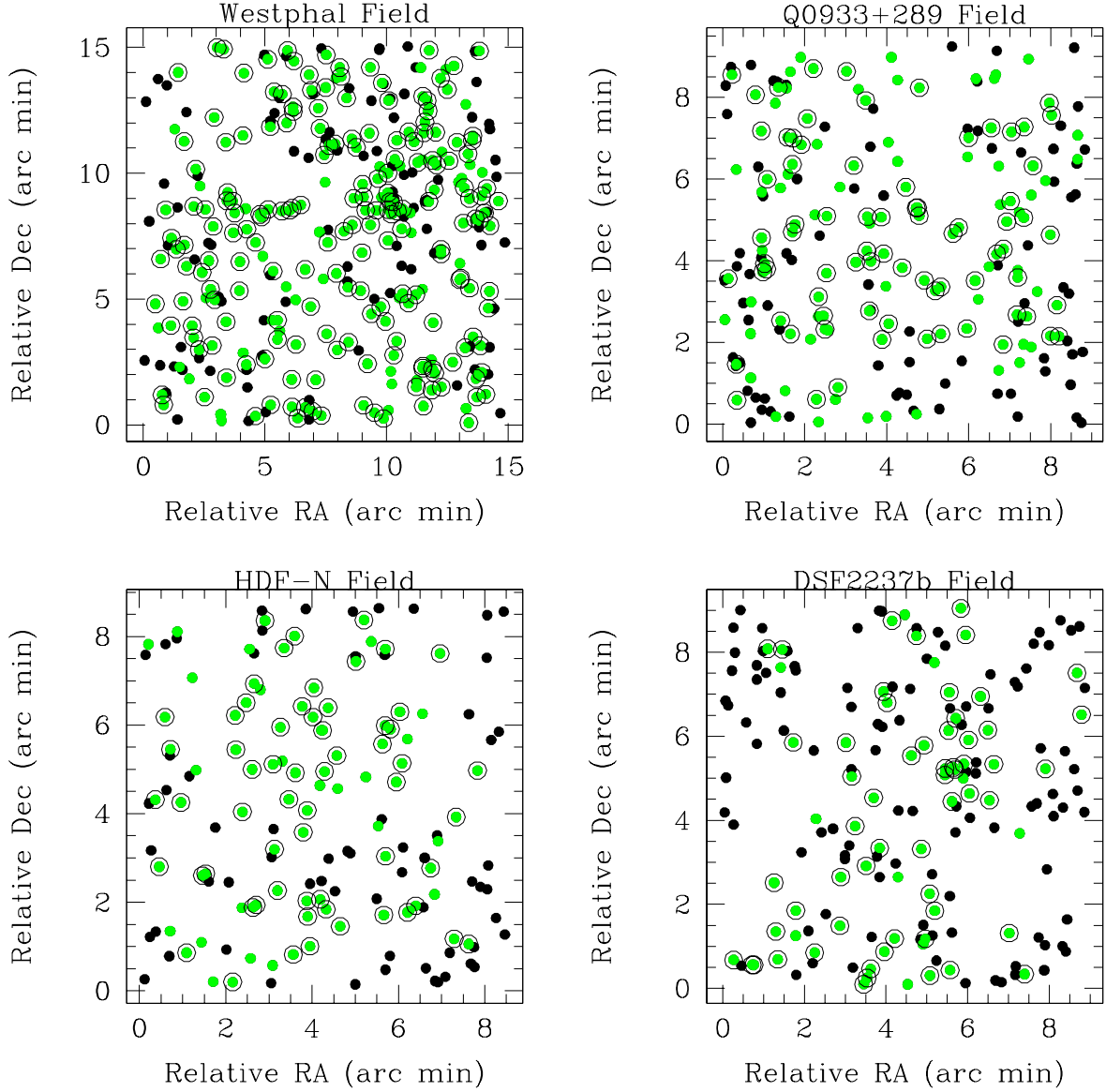}
\caption{Field maps of 4 of the survey fields; all dots are photometrically selected LBG
candidates. Green dots are objects that have been observed spectroscopically, and 
green dots with circles have spectroscopic redshifts $z > 2$. } 
\end{figure}
\begin{figure}
\plotone{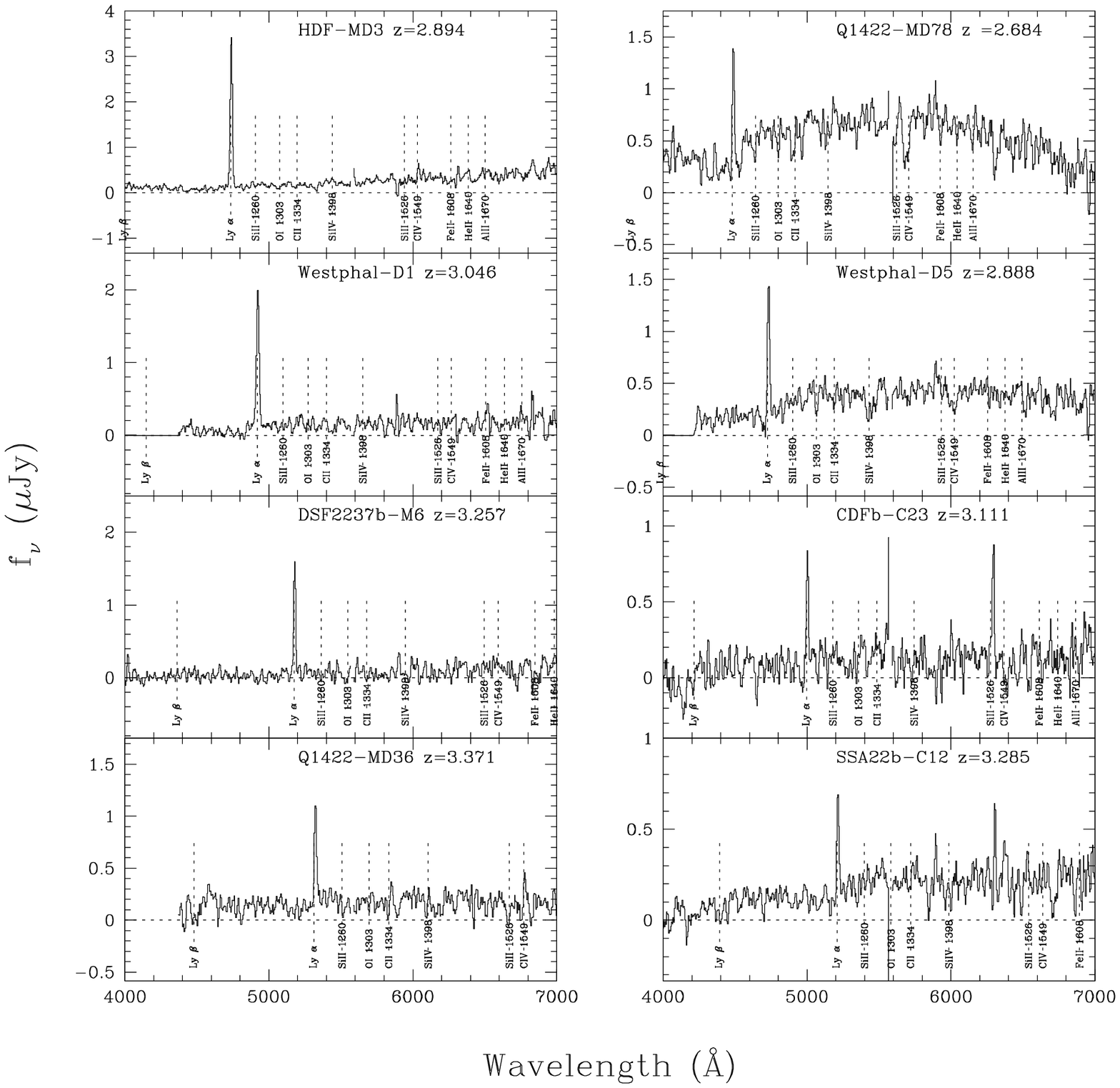}
\end{figure}
\clearpage
\begin{figure}
\plotone{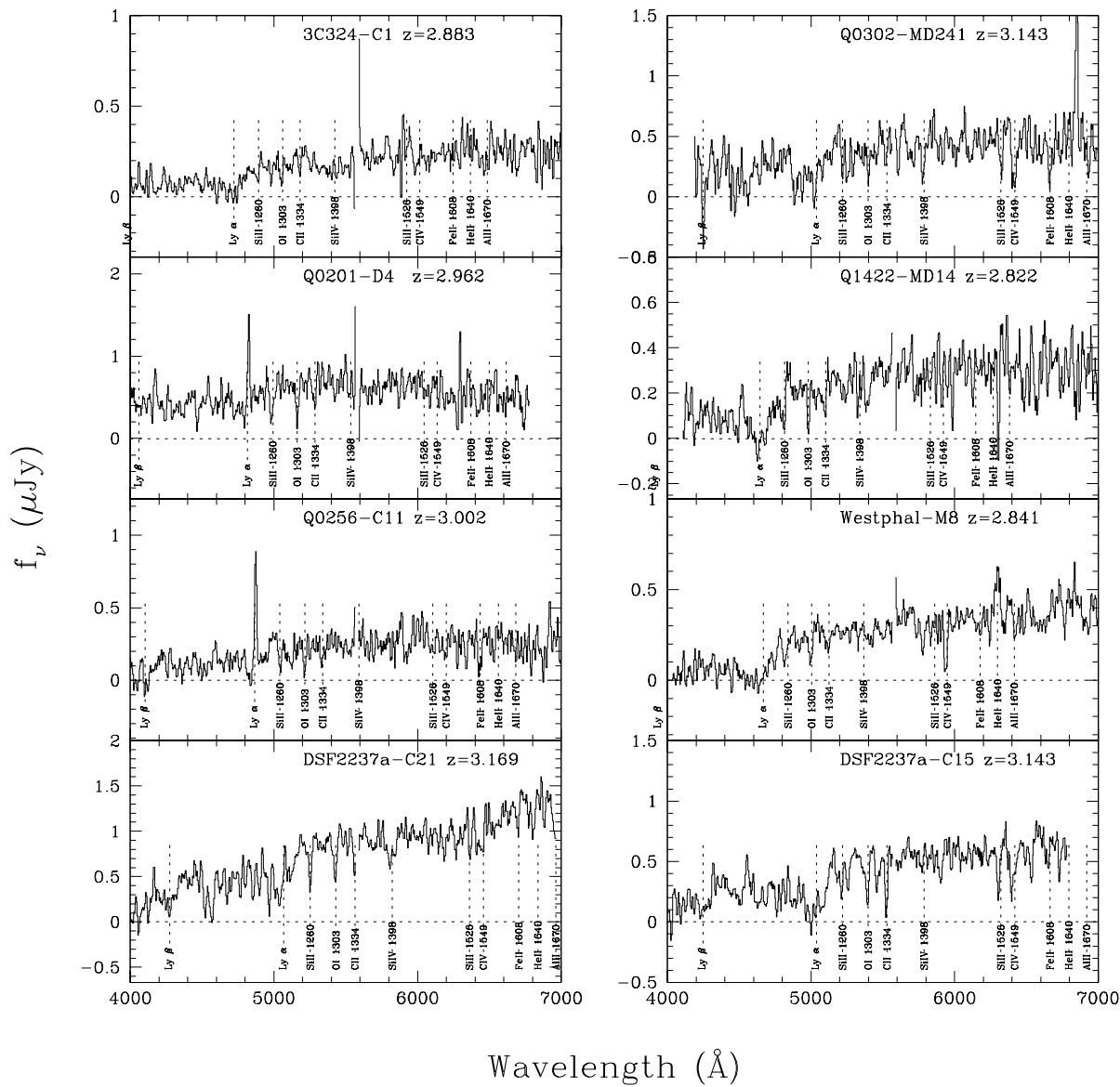}
\caption{Representative spectra of LBGs in the sample. These are intended to be typical
of spectra in the full sample, and to cover the range of spectral properties encountered.
They were chosen randomly from within lists of spectra grouped by the equivalent width of the
Lyman $\alpha$ feature. The spectra have been rebinned to 4.8 \AA\ per pixel (approximately
2 original pixels and half of a resolution element) and smoothed with a 3-pixel tapered box-car
for display purposes.}
\end{figure}
\begin{figure}
\plotone{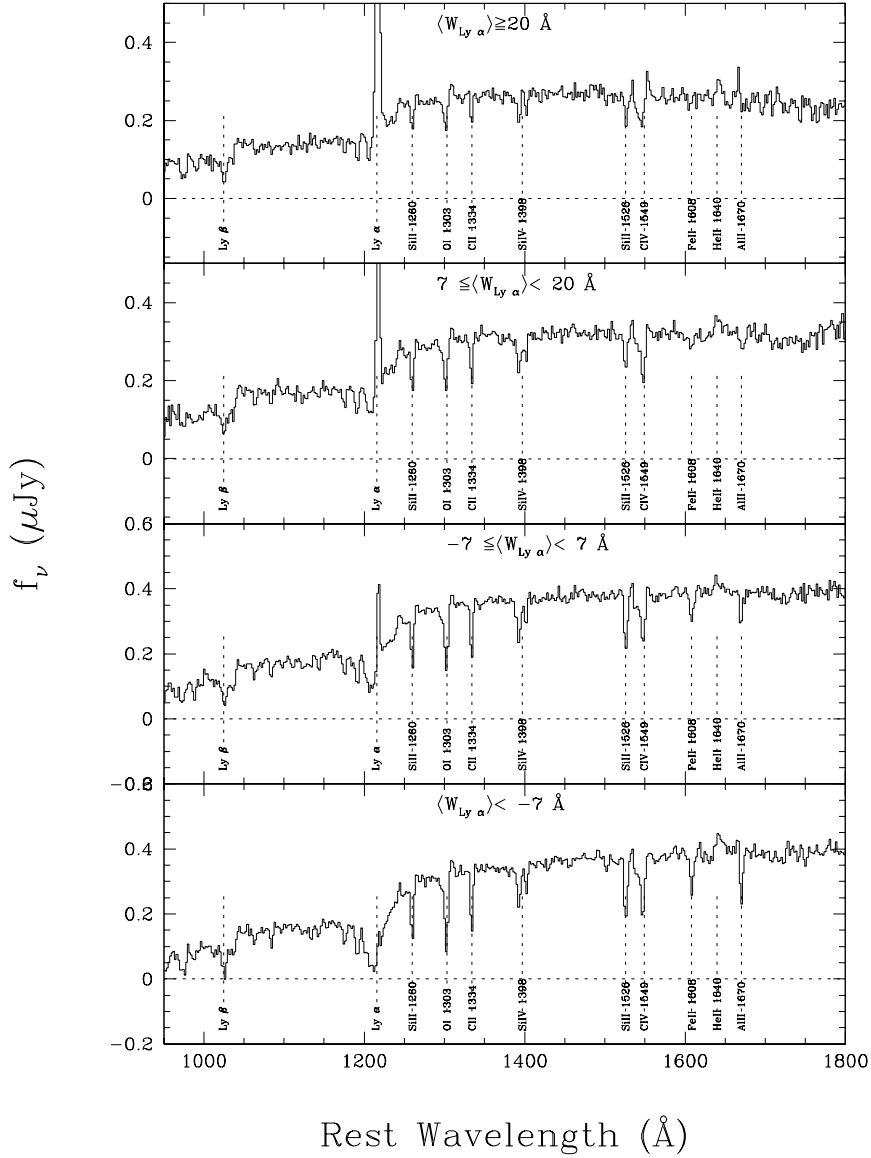}
\caption{Composite spectra of LBGs formed from $\sim 200$ spectra in each of 4 bins
in Lyman $\alpha$ emission strength. The plots have been scaled to emphasize the continuum features
in the spectra.  Significant trends discernible in the composite spectra are discussed in
Shapley \et 2003. }  
\end{figure}
\begin{figure}
\plotone{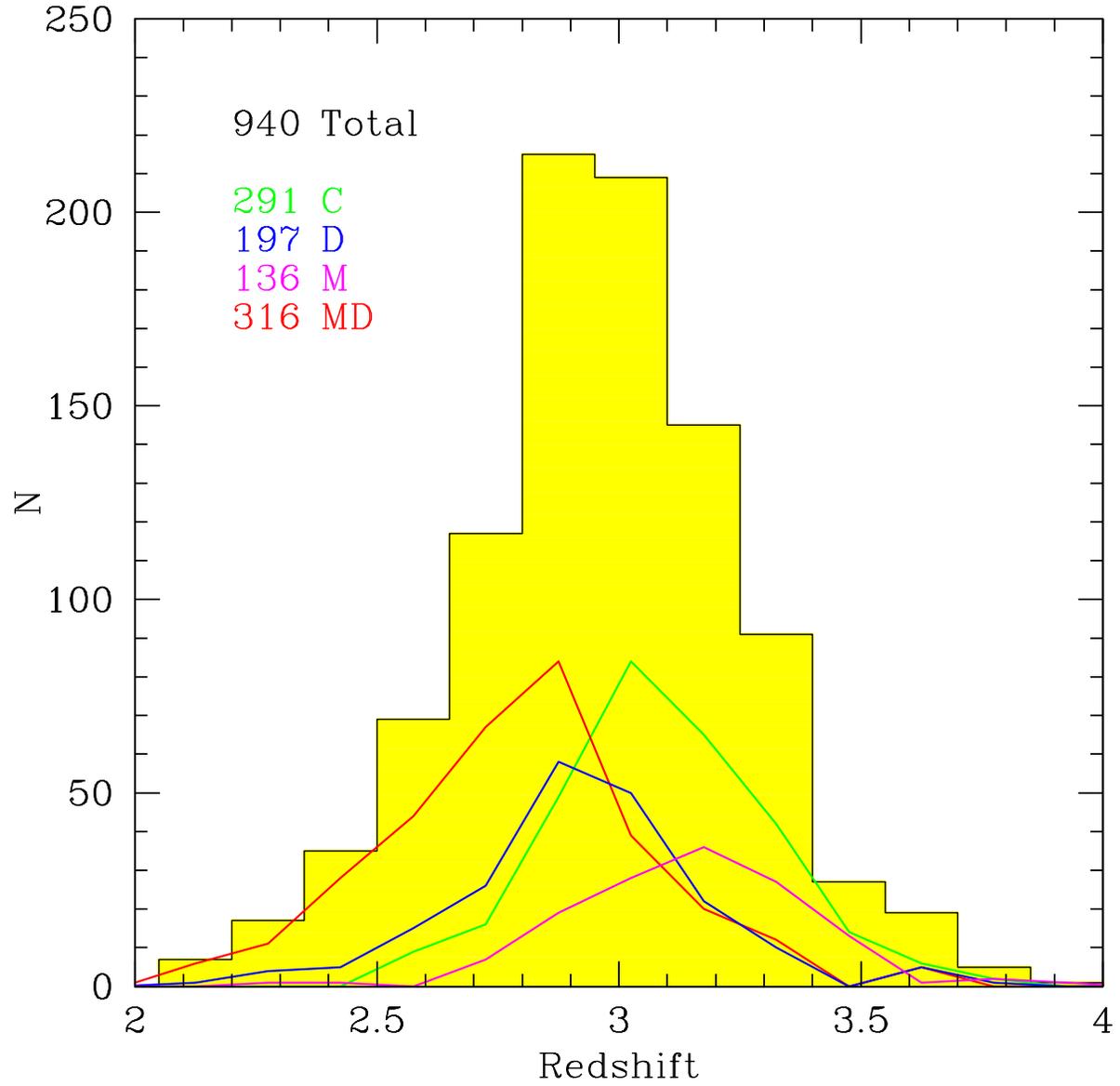}
\caption{Redshift histogram for the objects satisfying the the $z \sim 3$ LBG color
selection criteria (Table 4). Also shown are the histograms for each type of candidate
(C/D/M/MD). The statistics are summarized in Table 6. }
\end{figure}
\begin{figure}
\plottwo{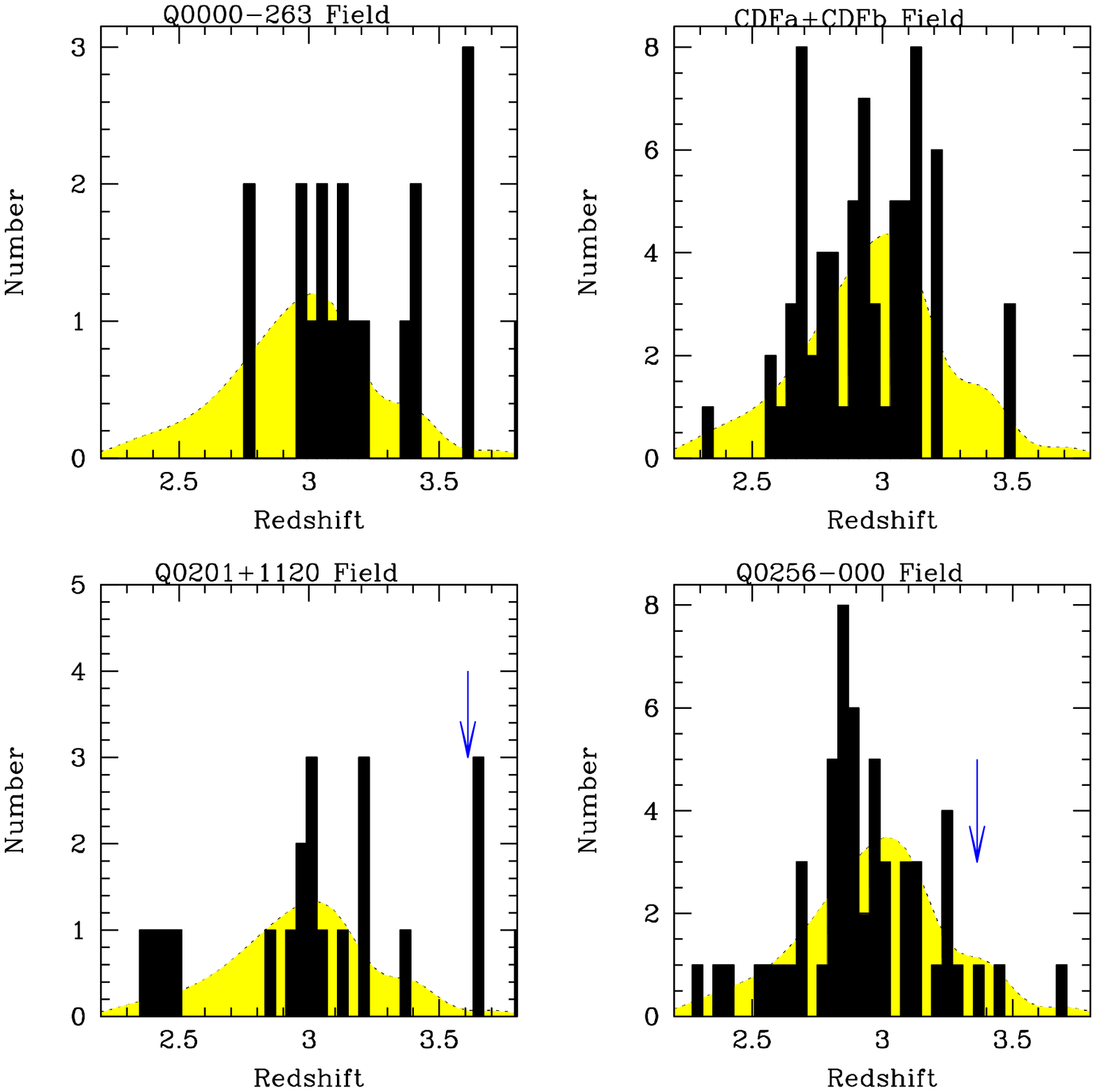}{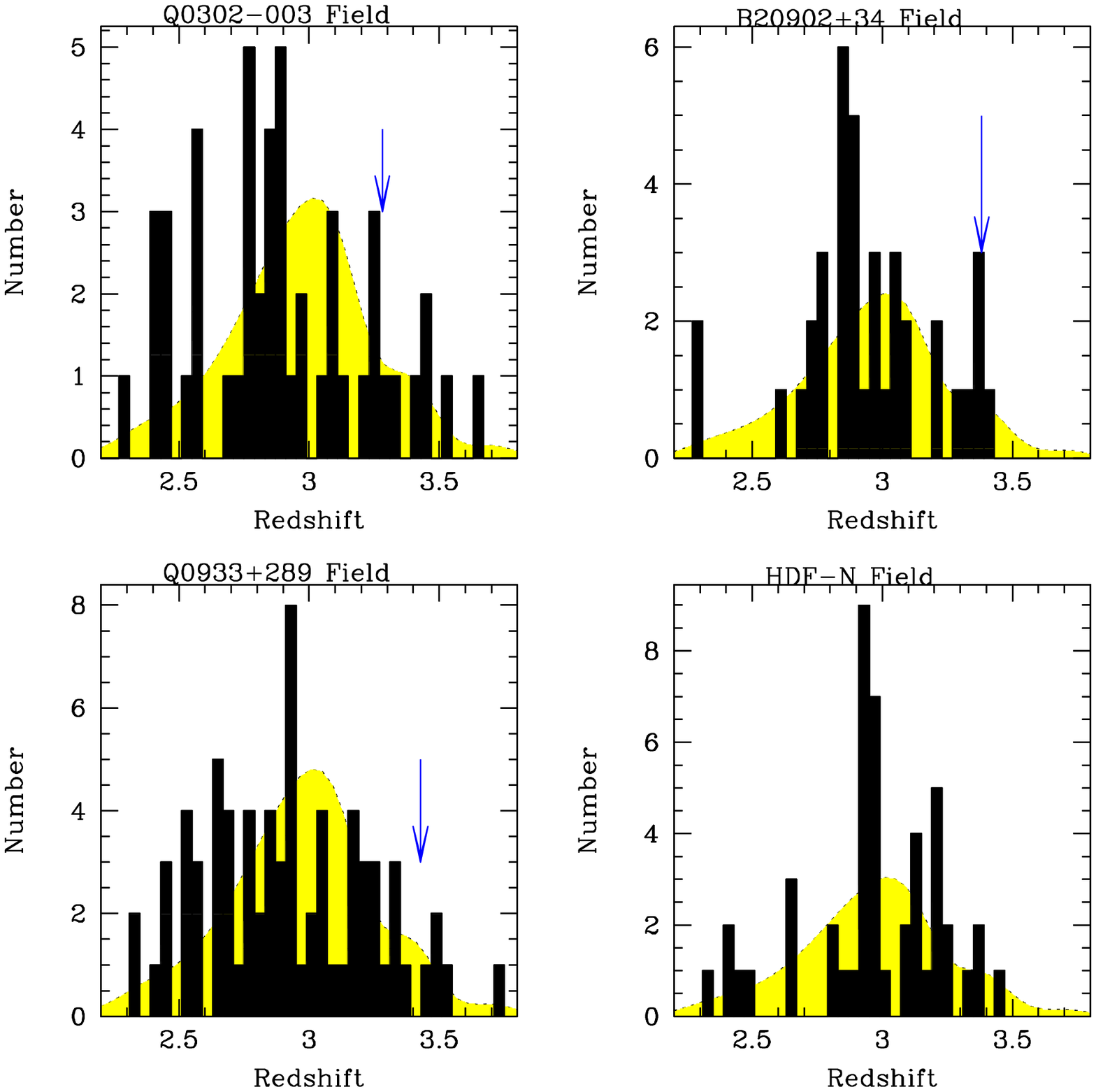}
\end{figure}
\begin{figure}
\plottwo{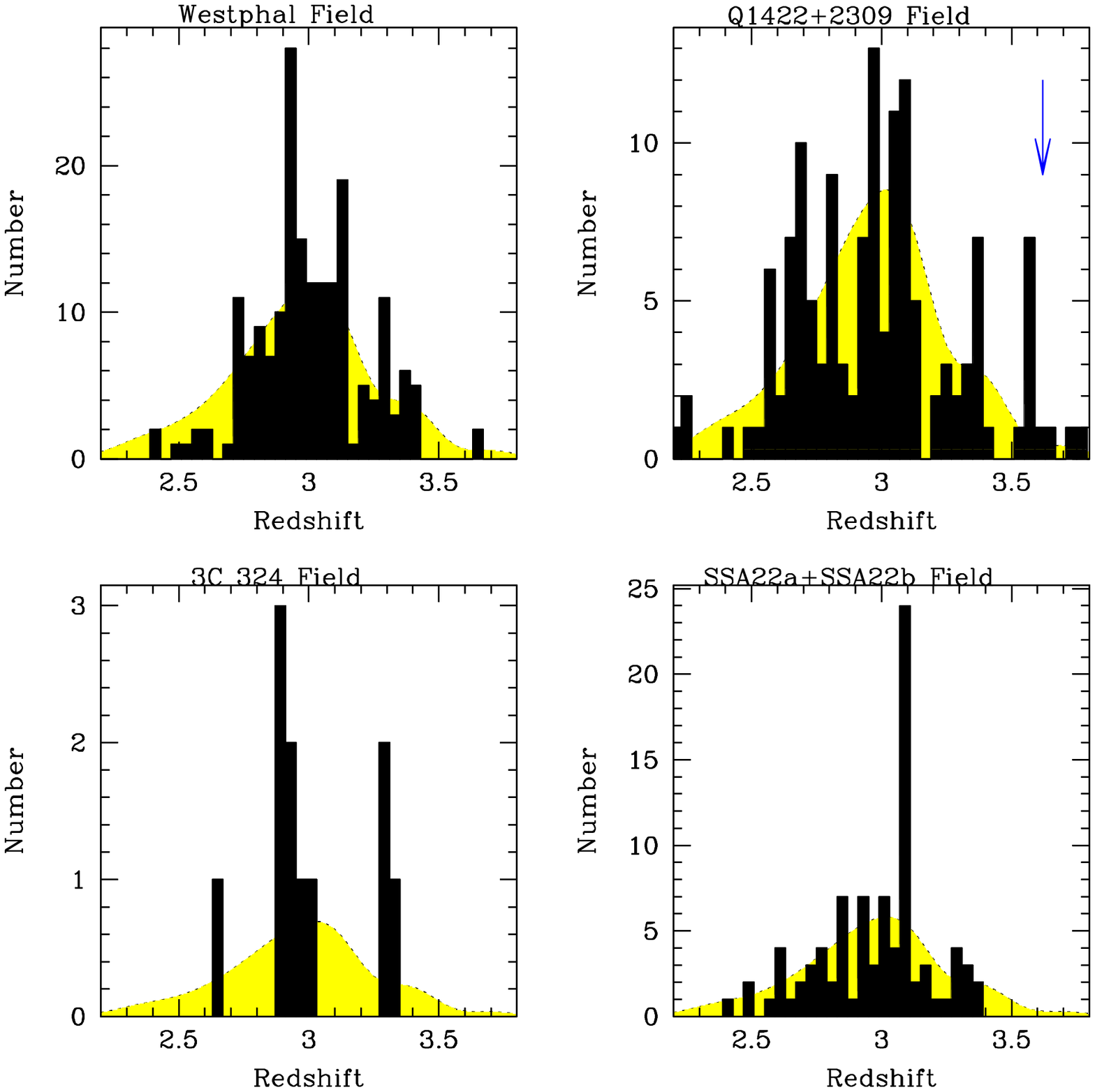}{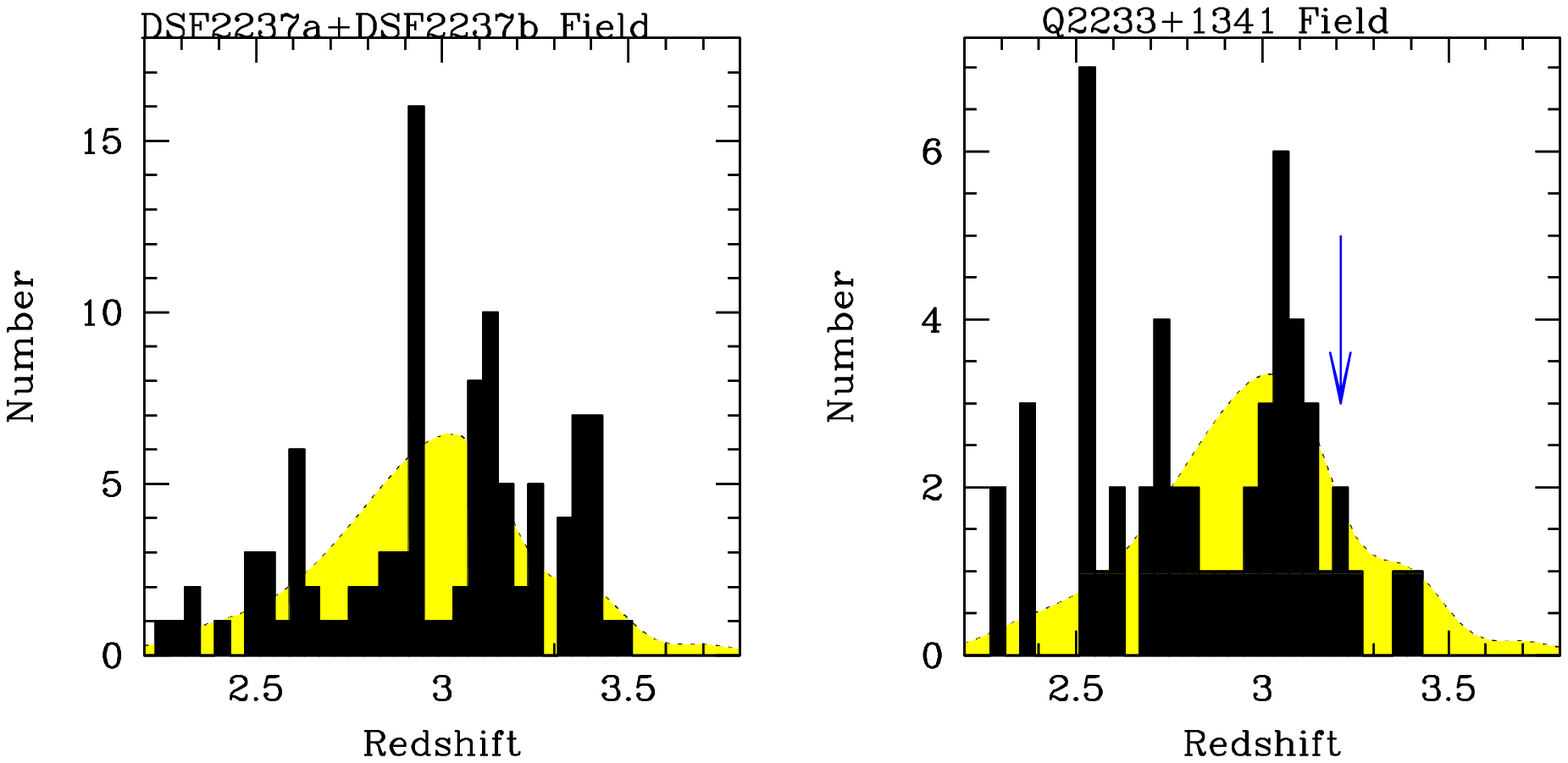}
\caption{Redshift histograms for individual fields, including only those objects
satisfying the LBG color criteria. The light colored histogram is the redshift
distribution for the full sample, normalized to the number of objects with redshifts
in each field. In cases where the survey fields were centered on the positions of known
QSOs or high redshift radio galaxies, an arrow marks the redshift for that object. In the
case of SSA22, DSF2237, and CDF, adjacent survey fields have been combined.}
\end{figure}
\begin{figure}
\plotone{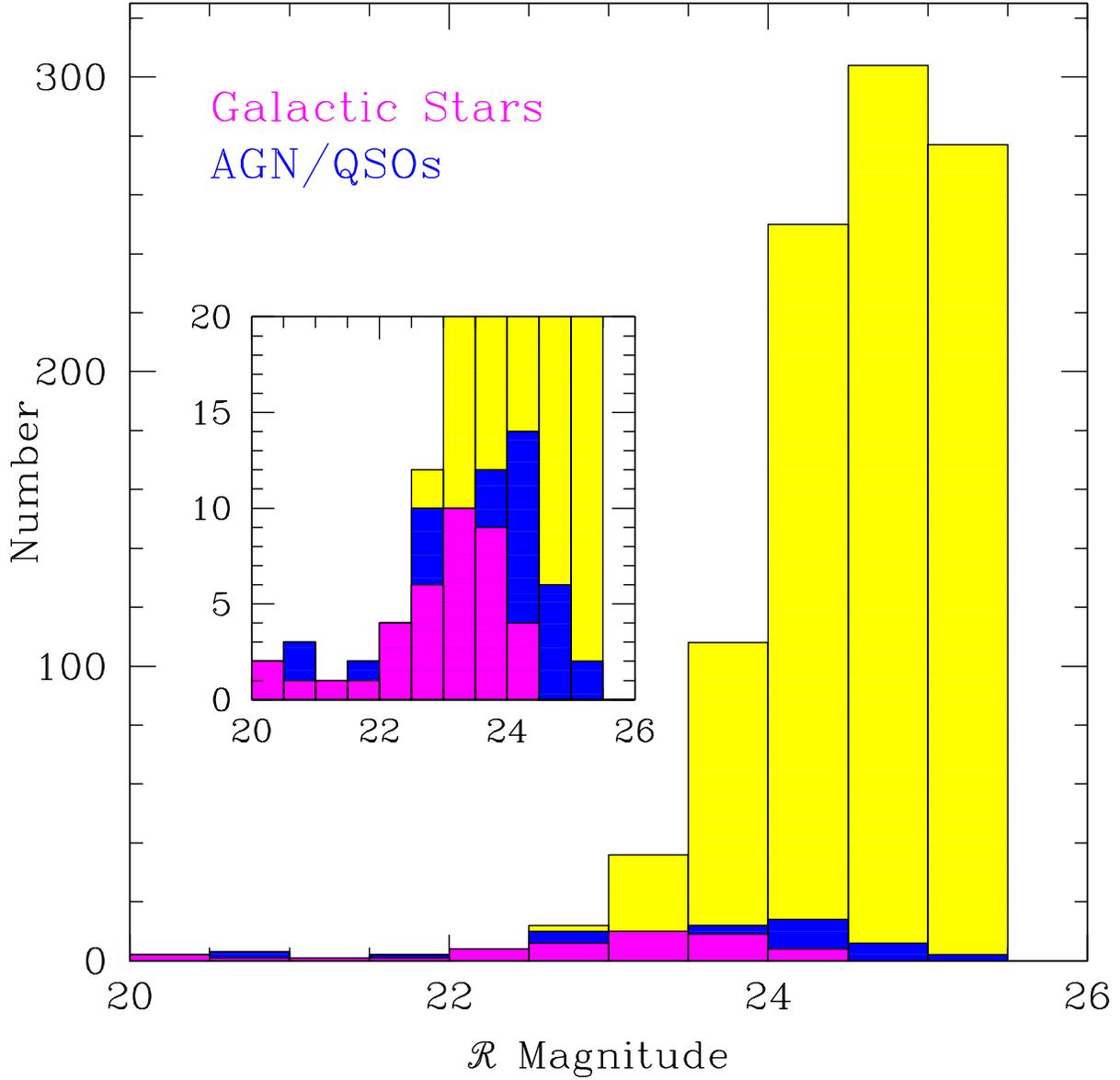}
\caption{Histogram of the apparent magnitude distribution of the contamination of
the LBG sample by stars (magenta) and AGN/QSOs (blue; added to the stellar histogram
so that the height of stars + AGN/QSOs is the total contamination of the sample) relative to the entire
spectroscopically identified sample (yellow). Note that the bright end of the magnitude
distribution is dominated by stellar and AGN contamination, but by ${\cal R}=24$, there is
essentially no stellar contamination and the AGN contamination is only a few percent. } 
\end{figure}
\end{document}

%% file: tab8_short.tex
\begin{deluxetable}{lcccccrrcc}
\tabletypesize{\scriptsize}
\tablewidth{0pc}
\tablecaption{CDFa Field LBGs\tablenotemark{*}}
\tablehead{
\colhead{Name} & \colhead{$\alpha(J2000)$} & \colhead{$\delta(J2000)$} & \colhead{${\cal R}$} &
\colhead{$G-{\cal R}$} & \colhead{$U_n-G$} & \colhead{${\rm z_{em}}$} & \colhead{${\rm z_{abs}}$} &
\colhead{Type} & \colhead{Notes\tablenotemark{a}}} 
\startdata
CDFa-C1 & 00:53:34.73 & 12:30:30.6 & 23.53 & 0.69 & 3.24 & -2.000 & 3.113 & GAL &   \\
CDFa-C2 & 00:53:06.89 & 12:30:46.1 & 25.45 & 0.36 & 2.24 & -1.000 & -1.000 & --- &   \\
CDFa-C3 & 00:53:14.01 & 12:29:46.6 & 24.69 & 0.52 & 2.85 & -2.000 & -2.000 & --- &   \\
CDFa-C4 & 00:53:18.63 & 12:31:08.5 & 25.18 & 0.66 & 2.18 & -2.000 & :2.628 & GAL &  \\
CDFa-C5 & 00:53:36.02 & 12:31:42.6 & 24.61 & 0.87 & 2.37 & -2.000 & -2.000 & --- &   \\
CDFa-C6 & 00:53:39.85 & 12:31:45.9 & 25.43 & 0.59 & 2.13 & 2.971 & 2.953 & GAL &   \\
CDFa-C7 & 00:53:37.62 & 12:31:58.2 & 23.13 & 0.73 & 3.74 & -2.000 & 3.072 & GAL &   \\
... & ... & ... & ... & ... & ...& ... & ... & ... & ... \\
\enddata
\tablenotetext{*}{The complete version of this table will be available in the electronic
version of the paper.}
\end{deluxetable}

%% file: tab9_short.tex
\begin{deluxetable}{lcccccrrcc}
\tabletypesize{\scriptsize}
\tablewidth{0pc}
\tablecaption{CDFb Field LBGs\tablenotemark{*}}
\tablehead{
\colhead{Name} & \colhead{$\alpha(J2000)$} & \colhead{$\delta(J2000)$} & \colhead{${\cal R}$} &
\colhead{$G-{\cal R}$} & \colhead{$U_n-G$} & \colhead{${\rm z_{em}}$} & \colhead{${\rm z_{abs}}$} &
\colhead{Type} & \colhead{Notes\tablenotemark{a}}} 
\startdata
CDFb-C1 & 00:53:38.95 & 12:20:54.9 & 25.37 & 0.55 & 2.09 & -1.000 & -1.000 & --- &   \\
CDFb-C2 & 00:54:00.01 & 12:21:36.3 & 25.02 & 0.60 & 2.18 & -1.000 & -1.000 & --- &   \\
CDFb-C3 & 00:53:35.71 & 12:21:30.7 & 24.18 & 0.84 & 2.73 & -2.000 & 3.477 & GAL &   \\
CDFb-C4 & 00:53:48.66 & 12:22:08.1 & 25.05 & 0.45 & 2.34 & -2.000 & -2.000 & --- &   \\
CDFb-C5 & 00:53:33.71 & 12:22:09.2 & 25.46 & 0.51 & 2.23 & -2.000 & -2.000 & --- &   \\
CDFb-C6 & 00:53:54.46 & 12:22:23.3 & 24.59 & 0.69 & 2.41 & -1.000 & -1.000 & --- &   \\
CDFb-C7 & 00:53:50.81 & 12:22:22.2 & 24.61 & 0.76 & 2.26 & -1.000 & -1.000 & --- &   \\
... & ... & ... & ... & ...& ... & ... & ... & ... & ... \\
\enddata
\tablenotetext{*}{The complete version of this table will be available in the electronic version
of the paper.}
\end{deluxetable}

%% file: tab10_short.tex
\begin{deluxetable}{lcccccrrcc}
\tabletypesize{\scriptsize}
\tablewidth{0pc}
\tablecaption{Q0201+1120 Field LBGs\tablenotemark{*}}
\tablehead{
\colhead{Name} & \colhead{$\alpha(J2000)$} & \colhead{$\delta(J2000)$} & \colhead{${\cal R}$} &
\colhead{$G-{\cal R}$} & \colhead{$U_n-G$} & \colhead{${\rm z_{em}}$} & \colhead{${\rm z_{abs}}$} &
\colhead{Type} & \colhead{Notes\tablenotemark{a}}} 
\startdata
Q0201-C1 & 02:03:59.94 & 11:30:20.4 & 24.54 & 0.43 & 2.56 & -1.000 & -1.000 & --- &   \\
Q0201-C2 & 02:03:32.97 & 11:31:24.2 & 24.22 & 0.95 & 2.75 & -1.000 & -1.000 & --- &   \\
Q0201-C3 & 02:03:56.25 & 11:30:54.9 & 24.94 & 0.51 & 2.42 & 3.020 & -2.000 & GAL &   \\
Q0201-C4 & 02:03:30.91 & 11:31:53.0 & 24.25 & 0.61 & 2.40 & -1.000 & -1.000 & --- &   \\
Q0201-C5 & 02:03:54.34 & 11:31:31.2 & 25.46 & 0.32 & 2.29 & -1.000 & -1.000 & --- &   \\
Q0201-C6 & 02:03:32.52 & 11:34:15.1 & 25.04 & 0.55 & 2.40 & -1.000 & -1.000 & --- &   \\
Q0201-C7 & 02:03:36.87 & 11:35:31.7 & 24.37 & 0.69 & 2.88 & 3.138 & 3.129 & GAL &  \\
... & ... & ... & ... & ...& ... & ... & ... & ... & ... \\
\enddata
\tablenotetext{*}{The complete version of this table will be available in the electronic version
of the paper.}
\tablenotetext{a}{Objects discussed in Ellison \et 2001.}
\tablenotetext{b}{This galaxy was previously known as Q0201-C6 in Pettini \et 1998,2001 and in Shapley
\et 2001.}   
\end{deluxetable}

%% file: tab11_short.tex
\begin{deluxetable}{lcccccrrcc}
\tabletypesize{\scriptsize}
\tablewidth{0pc}
\tablecaption{Q0256$-$000 Field LBGs\tablenotemark{*}}
\tablehead{
\colhead{Name} & \colhead{$\alpha(J2000)$} & \colhead{$\delta(J2000)$} & \colhead{${\cal R}$} &
\colhead{$G-{\cal R}$} & \colhead{$U_n-G$} & \colhead{${\rm z_{em}}$} & \colhead{${\rm z_{abs}}$} &
\colhead{Type} & \colhead{Notes\tablenotemark{a}}} 
\startdata
Q0256-C8 & 02:58:57.85 & 00:07:58.0 & 24.91 & 0.32 & 2.98 & -2.000 & -2.000 & --- &   \\
Q0256-C9 & 02:58:56.97 & 00:08:50.8 & 24.13 & 0.74 & 2.86 & -2.000 & -2.000 & --- &   \\
Q0256-C10 & 02:58:52.51 & 00:09:05.6 & 25.07 & 0.63 & 2.25 & -1.000 & -1.000 & --- &   \\
Q0256-C11 & 02:59:19.50 & 00:09:15.4 & 25.15 & 0.50 & 2.65 & 3.009 & 3.001 & GAL &   \\
Q0256-C14 & 02:58:53.07 & 00:09:53.6 & 24.71 & 0.48 & 2.91 & :2.983 & -2.000 & GAL &  \\
Q0256-C15 & 02:58:53.18 & 00:10:40.0 & 25.14 & 0.67 & 2.20 & -2.000 & 2.964 & GAL &   \\
Q0256-C16 & 02:58:53.11 & 00:10:43.4 & 24.22 & 0.98 & 2.83 & -2.000 & 2.974 & GAL &  \\
... & ... & ... & ... & ...& ... & ... & ... & ... & ... \\
\enddata
\tablenotetext{*}{The complete version of this table will be available in the electronic version
of the paper.}
\tablenotetext{a}{Objects observed in the near-IR are indicated
with their old designations as in Shapley \et 2001 (for cross-referencing).}
\end{deluxetable}

%% file: tab12_short.tex
\begin{deluxetable}{lcccccrrcc}
\tabletypesize{\scriptsize}
\tablewidth{0pc}
\tablecaption{Q0302$-$003 Field LBGs\tablenotemark{*}}
\tablehead{
\colhead{Name} & \colhead{$\alpha(J2000)$} & \colhead{$\delta(J2000)$} & \colhead{${\cal R}$} &
\colhead{$G-{\cal R}$} & \colhead{$U_n-G$} & \colhead{${\rm z_{em}}$} & \colhead{${\rm z_{abs}}$} &
\colhead{Type} & \colhead{Notes}}
\startdata
Q0302-C131 & 03:04:35.04 & -00:11:18.3 & 24.48 & 0.54 & 2.57 & 3.240 & 3.227 & GAL &   \\
Q0302-C133 & 03:04:33.17 & -00:11:46.9 & 24.55 & 0.84 & 2.37 & -1.000 & -1.000 & --- &   \\
Q0302-C134 & 03:04:29.20 & -00:11:00.3 & 25.18 & 0.37 & 2.41 & -1.000 & -1.000 & --- &   \\
Q0302-C140 & 03:04:52.11 & -00:10:21.7 & 25.28 & 0.48 & 2.16 & -1.000 & -1.000 & --- &   \\
Q0302-C141 & 03:04:53.42 & -00:10:17.1 & 24.33 & 0.56 & 2.48 & -2.000 & -2.000 & --- &   \\
Q0302-C146 & 03:04:26.45 & -00:10:06.9 & 25.21 & 0.50 & 2.48 & -1.000 & -1.000 & --- &   \\
Q0302-C148 & 03:04:33.59 & -00:09:51.2 & 24.27 & 0.51 & 2.67 & :2.848 & -2.000 & GAL &  \\
... & ... & ... & ... & ...& ... & ... & ... & ... & ... \\
\enddata
\tablenotetext{*}{The complete version of this table will be available in the electronic version
of the paper.}
\end{deluxetable}

%% file: tab13_short.tex
\begin{deluxetable}{lcccccrrcc}
\tabletypesize{\scriptsize}
\tablewidth{0pc}
\tablecaption{B20902$+$34 Field LBGs\tablenotemark{*}}
\tablehead{
\colhead{Name} & \colhead{$\alpha(J2000)$} & \colhead{$\delta(J2000)$} & \colhead{${\cal R}$} &
\colhead{$G-{\cal R}$} & \colhead{$U_n-G$} & \colhead{${\rm z_{em}}$} & \colhead{${\rm z_{abs}}$} &
\colhead{Type} & \colhead{Notes\tablenotemark{a}}} 
\startdata
B20902-C1 & 09:05:25.05 & 34:04:54.4 & 24.41 & 0.92 & 2.76 & -1.000 & -1.000 & --- &   \\
B20902-C2 & 09:05:37.86 & 34:05:57.9 & 25.16 & 0.56 & 2.48 & -2.000 & 2.732 & GAL &   \\
B20902-C4 & 09:05:45.78 & 34:08:18.0 & 24.52 & 0.63 & 2.70 & 3.034 & 3.025 & GAL &   \\
B20902-C5 & 09:05:23.09 & 34:08:59.3 & 24.70 & 0.93 & 2.47 & -2.000 & 3.098 & GAL &   \\
B20902-C6 & 09:05:20.58 & 34:09:07.7 & 24.13 & 0.45 & 3.45 & 3.099 & -2.000 & GAL &  \\
B20902-C7 & 09:05:30.11 & 34:09:07.7 & 24.52 & 0.37 & 3.16 & 3.195 & -2.000 & GAL &   \\
... & ... & ... & ... & ...& ... & ... & ... & ... & ... \\
\enddata
\tablenotetext{*}{The complete version of this table will be available in the electronic version
of the paper.}
\end{deluxetable}

%% file: tab14_short.tex
\begin{deluxetable}{lcccccrrcc}
\tabletypesize{\scriptsize}
\tablewidth{0pc}
\tablecaption{Q0933$+$289 Field LBGs\tablenotemark{*}}
\tablehead{
\colhead{Name} & \colhead{$\alpha(J2000)$} & \colhead{$\delta(J2000)$} & \colhead{${\cal R}$} &
\colhead{$G-{\cal R}$} & \colhead{$U_n-G$} & \colhead{${\rm z_{em}}$} & \colhead{${\rm z_{abs}}$} &
\colhead{Type} & \colhead{Notes\tablenotemark{a}}} 
\startdata
Q0933-C1 & 09:33:20.22 & 28:41:16.4 & 25.02 & 0.77 & 2.32 & -1.000 & -1.000 & --- &   \\
Q0933-C2 & 09:33:37.09 & 28:41:15.6 & 24.87 & 1.03 & 2.64 & -1.000 & -1.000 & --- &   \\
Q0933-C3 & 09:33:19.53 & 28:41:34.1 & 25.42 & 0.38 & 2.14 & -1.000 & -1.000 & --- &   \\
Q0933-C4 & 09:33:35.18 & 28:41:38.1 & 24.79 & 0.77 & 2.43 & -1.000 & -1.000 & --- &   \\
Q0933-C5 & 09:33:47.89 & 28:41:41.4 & 24.44 & 0.46 & 2.80 & -1.000 & -1.000 & --- &   \\
Q0933-C6 & 09:33:35.47 & 28:41:42.0 & 25.09 & 0.55 & 2.38 & -1.000 & -1.000 & --- &   \\
Q0933-C7 & 09:33:28.64 & 28:41:49.5 & 24.99 & 0.70 & 2.41 & :3.383 & :3.376 & GAL &  \\
... & ... & ... & ... & ...& ... & ... & ... & ... & ... \\
\enddata
\tablenotetext{*}{The complete version of this table will be available in the electronic version
of the paper.}
\end{deluxetable}

%% file: tab15_short.tex
\begin{deluxetable}{lcccccrrcc}
\tabletypesize{\scriptsize}
\tablewidth{0pc}
\tablecaption{HDF-N Field LBGs\tablenotemark{*}}
\tablehead{
\colhead{Name} & \colhead{$\alpha(J2000)$} & \colhead{$\delta(J2000)$} & \colhead{${\cal R}$} &
\colhead{$G-{\cal R}$} & \colhead{$U_n-G$} & \colhead{${\rm z_{em}}$} & \colhead{${\rm z_{abs}}$} &
\colhead{Type} & \colhead{Notes}} 
\startdata
HDF-C2 & 12:36:57.36 & 62:09:00.4 & 24.14 & 0.85 & 2.46 & -1.000 & -1.000 & --- & \\
HDF-C3 & 12:36:29.13 & 62:09:03.6 & 23.50 & 1.02 & 2.52 & -2.000 & -2.000 & --- & \\
HDF-C4 & 12:37:03.34 & 62:09:20.4 & 25.16 & 0.41 & 2.02 & -1.000 & -1.000 & --- & \\
HDF-C5 & 12:36:23.87 & 62:09:42.7 & 24.88 & 0.42 & 2.23 & -2.000 & :2.664 & GAL & \\
HDF-C6 & 12:37:16.96 & 62:10:02.3 & 24.34 & 0.55 & 2.74 & 3.451 & -2.000 & GAL & \\
HDF-C7 & 12:36:37.66 & 62:10:47.8 & 24.57 & 0.62 & 2.46 & -2.000 & 2.658 & GAL & \\
HDF-C8 & 12:36:26.95 & 62:11:27.4 & 24.38 & 0.70 & 2.35 & 2.993 & 2.983 & GAL & \\
HDF-C9 & 12:37:11.12 & 62:11:51.9 & 25.26 & 0.35 & 2.04 & -1.000 & -1.000 & --- & \\
HDF-C10 & 12:36:55.77 & 62:12:01.1 & 24.36 & 0.53 & 2.35 & -1.000 & -1.000 & --- & \\
HDF-C11 & 12:36:41.26 & 62:12:03.4 & 24.41 & 0.86 & 2.50 & 3.222 & 3.214 & GAL & C4-09\tablenotemark{a},4-858.0\tablenotemark{b} \\
... & ... & ... & ... & ...& ... & ... & ... & ... & ... \\
\enddata
\tablenotetext{*}{The complete version of this table will be available in the electronic version
of the paper.}
\tablenotetext{a}{Object within central HDF region observed with HST; designation as published in Steidel \et 1996b.}
\tablenotetext{b}{Object within central HDF WFPC-2 region; designation as published in Williams \et 1996.}
\tablenotetext{c}{Redshift from Lowenthal \et 1997.}
\end{deluxetable}

%% file: tab16_short.tex
\begin{deluxetable}{lcccccrrcc}
\tabletypesize{\scriptsize}
\tablewidth{0pc}
\tablecaption{Westphal Field LBGs\tablenotemark{*}}
\tablehead{
\colhead{Name\tablenotemark{a}} & \colhead{$\alpha(J2000)$} & \colhead{$\delta(J2000)$} & \colhead{${\cal R}$} &
\colhead{$G-{\cal R}$} & \colhead{$U_n-G$} & \colhead{${\rm z_{em}}$} & \colhead{${\rm z_{abs}}$} &
\colhead{Type} & \colhead{Notes\tablenotemark{a}}} 
\startdata
Westphal-C1 & 14:18:21.97 & 52:21:21.9 & 23.83 & 1.02 & 2.93 & 2.992 & 2.986 & GAL & CC1   \\
Westphal-C2 & 14:17:42.22 & 52:21:38.5 & 24.70 & 0.83 & 2.56 & 3.143 & 3.131 & GAL &   \\
Westphal-C3 & 14:17:37.88 & 52:21:57.3 & 25.21 & 0.54 & 2.32 & 2.896 & 2.888 & GAL &   \\
Westphal-C4 & 14:17:28.56 & 52:22:03.6 & 24.56 & 0.81 & 2.45 & -2.000 & :2.967 & GAL &  \\
Westphal-C5 & 14:18:24.15 & 52:22:23.4 & 24.81 & 0.67 & 2.50 & 2.815 & -2.000 & GAL &   \\
Westphal-C6 & 14:18:12.01 & 52:22:39.8 & 24.61 & 0.63 & 2.65 & 2.829 & 2.845 & GAL &   \\
Westphal-C7 & 14:18:01.21 & 52:22:53.9 & 25.27 & 0.71 & 2.23 & -2.000 & -2.000 & --- &   \\
... & ... & ... & ... & ...& ... & ... & ... & ... & ... \\
\enddata
\tablenotetext{*}{The complete version of this table will be available in the electronic version
of the paper.}
\tablenotetext{a}{For purely historical reasons, a number of objects in this field have been published
under slightly different names. For such cases, the previous designation is given in the ``Notes'' column.}
\end{deluxetable}

%% file: tab17_short.tex
\begin{deluxetable}{lcccccrrcc}
\tabletypesize{\scriptsize}
\tablewidth{0pc}
\tablecaption{3C 324 Field LBGs\tablenotemark{*}}
\tablehead{
\colhead{Name} & \colhead{$\alpha(J2000)$} & \colhead{$\delta(J2000)$} & \colhead{${\cal R}$} &
\colhead{$G-{\cal R}$} & \colhead{$U_n-G$} & \colhead{${\rm z_{em}}$} & \colhead{${\rm z_{abs}}$} &
\colhead{Type} & \colhead{Notes}} 
\startdata
3C324-C1 & 15:49:54.27 & 21:26:33.1 & 24.33 & 0.62 & 2.52 & 2.890 & 2.878 & GAL &   \\
3C324-C2 & 15:49:53.98 & 21:26:35.5 & 24.26 & 0.54 & 3.00 & -2.000 & 2.877 & GAL &   \\
3C324-C3 & 15:49:47.10 & 21:27:05.4 & 24.14 & 0.85 & 2.61 & 3.295 & 3.282 & GAL &   \\
3C324-C4 & 15:49:46.04 & 21:27:46.2 & 24.59 & 0.52 & 2.54 & -2.000 & 3.281 & GAL &   \\
3C324-C5 & 15:49:44.21 & 21:28:22.0 & 24.68 & 0.70 & 2.28 & -1.000 & -1.000 & --- &   \\
3C324-C6 & 15:49:38.74 & 21:29:17.5 & 24.73 & 0.83 & 2.33 & 3.313 & 3.303 & GAL &   \\
3C324-C7 & 15:49:47.47 & 21:29:21.1 & 25.09 & 0.61 & 2.42 & -1.000 & -1.000 & --- &   \\
... & ... & ... & ... & ...& ... & ... & ... & ... & ... \\
\enddata
\tablenotetext{*}{The complete version of this table will be available in the electronic version
of the paper.}
\end{deluxetable}

%% file: tab18_short.tex
\begin{deluxetable}{lcccccrrcc}
\tabletypesize{\scriptsize}
\tablewidth{0pc}
\tablecaption{Q1422$+$2309 Field LBGs\tablenotemark{*}}
\tablehead{
\colhead{Name} & \colhead{$\alpha(J2000)$} & \colhead{$\delta(J2000)$} & \colhead{${\cal R}$} &
\colhead{$G-{\cal R}$} & \colhead{$U_n-G$} & \colhead{${\rm z_{em}}$} & \colhead{${\rm z_{abs}}$} &
\colhead{Type} & \colhead{Notes}} 
\startdata
Q1422-C1 & 14:24:41.60 & 22:46:21.1 & 25.21 & 0.97 & 2.87 & -1.000 & -1.000 & --- &   \\
Q1422-C2 & 14:24:31.42 & 22:46:41.2 & 25.76 & 0.61 & 2.99 & -1.000 & -1.000 & --- &   \\
Q1422-C3 & 14:24:45.89 & 22:47:25.7 & 25.68 & 0.50 & 2.97 & -1.000 & -1.000 & --- &   \\
Q1422-C4 & 14:24:26.54 & 22:47:32.0 & 25.69 & 0.81 & 2.38 & -1.000 & -1.000 & --- &   \\
Q1422-C5 & 14:24:44.21 & 22:47:32.0 & 25.54 & 0.84 & 2.63 & -1.000 & -1.000 & --- &   \\
Q1422-C6 & 14:24:22.37 & 22:47:40.0 & 24.45 & 0.94 & 3.66 & -1.000 & -1.000 & --- &   \\
Q1422-C7 & 14:24:29.41 & 22:47:59.7 & 25.66 & 0.80 & 2.69 & -1.000 & -1.000 & --- &   \\
... & ... & ... & ... & ...& ... & ... & ... & ... & ... \\
\enddata
\tablenotetext{*}{The complete version of this table will be available in the electronic version
of the paper.}
\tablenotetext{a}{This object does not satisfy the LBG color selection criteria, but
is of interest because it is near Q1422+2309 on the plan of the sky and is at close to the
same redshift.}
\end{deluxetable}

%% file: tab19_short.tex
\begin{deluxetable}{lcccccrrcc}
\tabletypesize{\scriptsize}
\tablewidth{0pc}
\tablecaption{SSA22a Field LBGs\tablenotemark{*}}
\tablehead{
\colhead{Name\tablenotemark{a}} & \colhead{$\alpha(J2000)$} & \colhead{$\delta(J2000)$} & \colhead{${\cal R}$} &
\colhead{$G-{\cal R}$} & \colhead{$U_n-G$} & \colhead{${\rm z_{em}}$} & \colhead{${\rm z_{abs}}$} &
\colhead{Type} & \colhead{Notes}} 
\startdata
SSA22a-C1 & 22:17:31.36 & 00:10:41.7 & 25.34 & 0.55 & 2.17 & -1.000 & -1.000 & --- &   \\
SSA22a-C2 & 22:17:44.52 & 00:10:57.6 & 25.31 & 0.46 & 2.07 & -1.000 & -1.000 & --- &  \\
SSA22a-C3 & 22:17:32.53 & 00:10:57.4 & 25.05 & 0.35 & 2.36 & -1.000 & -1.000 & --- &  \\
SSA22a-C4 & 22:17:38.89 & 00:11:02.1 & 24.53 & 0.42 & 2.54 & 3.076 & -2.000 & GAL & S98-C3 \\
SSA22a-C6 & 22:17:40.93 & 00:11:26.0 & 23.44 & 0.79 & 2.97 & 3.098 & 3.092 & GAL & S98-C5 \\
SSA22a-C7 & 22:17:24.59 & 00:11:31.3 & 24.13 & 1.06 & 2.70 & -1.000 & -1.000 & --- &   \\
SSA22a-C8 & 22:17:27.64 & 00:11:59.0 & 25.38 & 0.20 & 2.22 & -1.000 & -1.000 & --- &  \\
... & ... & ... & ... & ...& ... & ... & ... & ... & ... \\
\enddata
\tablenotetext{*}{The complete version of this table will be available in the electronic version
of the paper.}
\tablenotetext{a}{Objects whose photometry satisfies our LBG selection criteria, but which
have been discussed under alternative names, are indicated in the ``Notes'' column. ``S96''
refers to Steidel \et 1996a, and ``S98'' refers to Steidel \et 1998. Objects which at one time
satisfied our selection criteria, and which were spectroscopically observed, are included
at the end of the table (see appendix).}
\end{deluxetable}

%% file: tab20_short.tex
\begin{deluxetable}{lcccccrrcc}
\tabletypesize{\scriptsize}
\tablewidth{0pc}
\tablecaption{SSA22b Field LBGs\tablenotemark{*}}
\tablehead{
\colhead{Name} & \colhead{$\alpha(J2000)$} & \colhead{$\delta(J2000)$} & \colhead{${\cal R}$} &
\colhead{$G-{\cal R}$} & \colhead{$U_n-G$} & \colhead{${\rm z_{em}}$} & \colhead{${\rm z_{abs}}$} &
\colhead{Type} & \colhead{Notes}} 
\startdata
SSA22b-C1 & 22:17:46.08 & 00:02:12.5 & 24.62 & 0.79 & 2.81 & -1.000 & -1.000 & --- &  \\
SSA22b-C3 & 22:17:29.22 & 00:02:48.8 & 24.55 & 0.68 & 2.68 & 2.917 & 2.911 & GAL &  \\
SSA22b-C4 & 22:17:46.85 & 00:03:29.2 & 25.47 & 0.32 & 2.07 & -2.000 & -2.000 & --- &  \\
SSA22b-C5 & 22:17:47.06 & 00:04:25.7 & 25.16 & 0.71 & 2.48 & 3.117 & 3.110 & GAL &   \\
SSA22b-C6 & 22:17:33.88 & 00:05:15.5 & 25.35 & 0.41 & 2.21 & 3.313 & -2.000 & GAL &   \\
SSA22b-C9 & 22:17:22.42 & 00:06:47.6 & 24.69 & 0.72 & 2.30 & 3.175 & 3.170 & GAL &  \\
SSA22b-C10 & 22:17:25.46 & 00:06:59.2 & 24.11 & 0.93 & 2.62 & 3.360 & 3.353 & GAL &  \\
... & ... & ... & ... & ...& ... & ... & ... & ... & ... \\
\enddata
\tablenotetext{*}{The complete version of this table will be available in the electronic version
of the paper.}
\tablenotetext{a}{This object appeared as a member of the $z=3.09$ ``spike'' in Steidel \et 1998,
called ``SSA22b-D27''.}
\end{deluxetable}

%% file: tab21_short.tex
\begin{deluxetable}{lcccccrrcc}
\tabletypesize{\scriptsize}
\tablewidth{0pc}
\tablecaption{DSF2237a Field LBGs\tablenotemark{*}}
\tablehead{
\colhead{Name} & \colhead{$\alpha(J2000)$} & \colhead{$\delta(J2000)$} & \colhead{${\cal R}$} &
\colhead{$G-{\cal R}$} & \colhead{$U_n-G$} & \colhead{${\rm z_{em}}$} & \colhead{${\rm z_{abs}}$} &
\colhead{Type} & \colhead{Notes}} 
\startdata
DSF2237a-C1 & 22:40:15.74 & 11:48:14.8 & 24.80 & 0.29 & 2.83 & -1.000 & -1.000 & --- &   \\
DSF2237a-C2 & 22:40:08.29 & 11:49:04.8 & 23.55 & 1.13 & 3.20 & 3.331 & 3.318 & GAL &   \\
DSF2237a-C3 & 22:40:23.15 & 11:49:28.6 & 24.48 & 0.75 & 2.64 & 3.111 & 3.107 & GAL &   \\
DSF2237a-C4 & 22:40:06.48 & 11:49:29.9 & 24.58 & 0.69 & 2.75 & 3.487 & 3.474 & GAL &   \\
DSF2237a-C5 & 22:40:22.88 & 11:49:38.7 & 25.02 & 0.60 & 2.28 & -1.000 & -1.000 & --- &   \\
DSF2237a-C6 & 22:40:08.26 & 11:49:45.6 & 24.26 & 1.04 & 2.97 & -2.000 & 3.145 & GAL &   \\
DSF2237a-C7 & 22:40:20.90 & 11:50:21.8 & 25.16 & 0.50 & 2.46 & 3.202 & 3.195 & GAL &   \\
... & ... & ... & ... & ...& ... & ... & ... & ... & ... \\
\enddata
\tablenotetext{*}{The complete version of this table will be available in the electronic version
of the paper.}
\end{deluxetable}

%% file: tab22_short.tex
\begin{deluxetable}{lcccccrrcc}
\tabletypesize{\scriptsize}
\tablewidth{0pc}
\tablecaption{DSF2237b Field LBGs\tablenotemark{*}}
\tablehead{
\colhead{Name} & \colhead{$\alpha(J2000)$} & \colhead{$\delta(J2000)$} & \colhead{${\cal R}$} &
\colhead{$G-{\cal R}$} & \colhead{$U_n-G$} & \colhead{${\rm z_{em}}$} & \colhead{${\rm z_{abs}}$} &
\colhead{Type} & \colhead{Notes}} 
\startdata
DSF2237b-C1 & 22:39:16.90 & 11:47:47.5 & 24.99 & 0.37 & 2.53 & 3.068 & 3.057 & GAL &   \\
DSF2237b-C2 & 22:39:31.96 & 11:47:58.8 & 24.95 & 0.85 & 2.40 & -2.000 & :3.142 & GAL &  \\
DSF2237b-C3 & 22:39:35.83 & 11:48:09.5 & 25.19 & 0.73 & 2.24 & -2.000 & -2.000 & --- &   \\
DSF2237b-C7 & 22:39:32.96 & 11:48:17.5 & 25.10 & 0.28 & 2.58 & -2.000 & :2.913 & GAL &  \\
DSF2237b-C8 & 22:39:36.05 & 11:48:17.1 & 24.93 & 0.73 & 2.27 & 3.107 & 3.110 & GAL &   \\
DSF2237b-C9 & 22:39:44.46 & 11:48:25.6 & 25.45 & 0.42 & 2.35 & 3.245 & 3.234 & GAL &   \\
DSF2237b-C10 & 22:39:24.33 & 11:48:28.5 & 25.10 & 0.60 & 2.55 & -1.000 & -1.000 & --- &   \\
... & ... & ... & ... & ...& ... & ... & ... & ... & ... \\
\enddata
\tablenotetext{*}{The complete version of this table will be available in the electronic version
of the paper.}
\end{deluxetable}

%% file: tab23_short.tex
\begin{deluxetable}{lcccccrrcc}
\tabletypesize{\scriptsize}
\tablewidth{0pc}
\tablecaption{Q2233$+$1341 Field LBGs\tablenotemark{*}}
\tablehead{
\colhead{Name} & \colhead{$\alpha(J2000)$} & \colhead{$\delta(J2000)$} & \colhead{${\cal R}$} &
\colhead{$G-{\cal R}$} & \colhead{$U_n-G$} & \colhead{${\rm z_{em}}$} & \colhead{${\rm z_{abs}}$} &
\colhead{Type} & \colhead{Notes}} 
\startdata
Q2233-C1 & 22:36:28.12 & 13:51:46.4 & 25.35 & 0.55 & 2.22 & -1.000 & -1.000 & --- &   \\
Q2233-C2 & 22:36:21.64 & 13:52:16.3 & 24.20 & 0.90 & 2.49 & 2.970 & 2.964 & GAL &   \\
Q2233-C3 & 22:36:22.97 & 13:52:33.9 & 24.61 & 0.66 & 2.48 & 2.546 & 2.542 & GAL &   \\
Q2233-C4 & 22:36:16.67 & 13:53:58.4 & 24.47 & 1.01 & 2.84 & 3.053 & -2.000 & GAL &   \\
Q2233-C5 & 22:36:44.87 & 13:54:30.7 & 25.32 & 0.42 & 2.08 & -1.000 & -1.000 & --- &   \\
Q2233-C6 & 22:36:22.55 & 13:54:29.7 & 24.72 & 0.59 & 2.09 & 2.546 & 2.542 & GAL &   \\
Q2233-C7 & 22:36:12.56 & 13:55:14.0 & 24.97 & 1.01 & 2.74 & -1.000 & -1.000 & --- &   \\
... & ... & ... & ... & ...& ... & ... & ... & ... & ... \\
\enddata
\tablenotetext{*}{The complete version of this table will be available in the electronic version
of the paper.}
\end{deluxetable}